\documentclass[10pt,conference]{IEEEtran}
\usepackage{cite}
\usepackage{amsmath,amssymb,amsfonts}
\usepackage{algorithmic}
\usepackage{graphicx}
\usepackage{textcomp}
\usepackage[dvipsnames]{xcolor}
\usepackage[hyphens]{url}
\usepackage{fancyhdr}
\usepackage{hyperref}

\usepackage{xspace}
\usepackage{hhline}
\usepackage{makecell}
\usepackage[utf8]{inputenc}
\usepackage{kotex} 
\usepackage{tcolorbox}
\usepackage{lipsum}
\usepackage{xfrac}
\usepackage{booktabs}
\usepackage{enumitem}
\usepackage{color,soul}
\usepackage{comment}
\usepackage{fancybox}
\usepackage{multirow}
\usepackage{balance}

\newcommand{\name}{$\texttt{RoMe}$\xspace}

\newcommand{\prefill}{$\texttt{prefill}$\xspace}
\newcommand{\decode}{$\texttt{decode}$\xspace}

\newcommand{\rdrow}{$\texttt{RD\_{row}}$\xspace}
\newcommand{\wrrow}{$\texttt{WR\_{row}}$\xspace}

\newcommand{\cba}{$\texttt{VBA}$\xspace}
\newcommand{\cbas}{$\texttt{VBA}$s\xspace}

\newcommand{\agbank}{$AG_{bank}$\xspace}
\newcommand{\agmc}{$AG_{MC}$\xspace}

\newcommand{\lbr}{$LBR$\xspace}
\newcommand{\lbrattn}{$LBR_{Attn}$\xspace}
\newcommand{\lbrffn}{$LBR_{FFN}$\xspace}

\newcommand{\trtrs}{$\texttt{tR2RS}$\xspace}
\newcommand{\trtrr}{$\texttt{tR2RR}$\xspace}
\newcommand{\trtws}{$\texttt{tR2WS}$\xspace}
\newcommand{\trtwr}{$\texttt{tR2WR}$\xspace}
\newcommand{\twtrs}{$\texttt{tW2RS}$\xspace}
\newcommand{\twtrr}{$\texttt{tW2RR}$\xspace}
\newcommand{\twtws}{$\texttt{tW2WS}$\xspace}
\newcommand{\twtwr}{$\texttt{tW2WR}$\xspace}
\newcommand{\trdrow}{$\texttt{tRD\_row}$\xspace}
\newcommand{\twrrow}{$\texttt{tWR\_row}$\xspace}
\newcommand{\trc}{$\texttt{tRC}$\xspace}
\newcommand{\trcdrd}{$\texttt{tRCDRD}$\xspace}

\newcommand{\tccdl}{$\texttt{tCCDL}$\xspace}
\newcommand{\tccds}{$\texttt{tCCDS}$\xspace}

\newcommand{\trrds}{$\texttt{tRRDS}$\xspace}

\newcommand{\trfcpb}{$\texttt{tRFCpb}$\xspace}
\newcommand{\trefi}{$\texttt{tREFI}$\xspace}
\newcommand{\trefipb}{$\texttt{tREFIpb}$\xspace}
\newcommand{\trtw}{$\texttt{tRTW}$\xspace}
\newcommand{\twtrss}{$\texttt{tWTRS}$\xspace}

\newcommand{\trrefd}{$\texttt{tRREFD}$\xspace}

\newcommand{\ie}{\textit{i.e.}\xspace}
\newcommand{\eg}{\textit{e.g.}\xspace}

\newcommand{\rowbuffer}{$\texttt{row-buffer}$\xspace}

\definecolor{lightergray}{gray}{0.95} 

\newcommand*{\Scale}[2][4]{\scalebox{#1}{$#2$}}

\newcommand{\hpcayear}{2026}

\title{RoMe: Row Granularity Access Memory System for Large Language Models}

\def\hpcacameraready{} 

\newcommand\hpcaauthors{Hwayong Nam$^\dagger$$^{*}$, Seungmin Baek$^\dagger$$^{*}$, Jumin Kim$^\dagger$, Michael Jaemin Kim$^\ddagger$, and Jung Ho Ahn$^\dagger$}
\newcommand\hpcaaffiliation{Seoul National University$^\dagger$, Meta$^\ddagger$}
\newcommand\hpcaemail{$^\dagger$\{nhy4916, qortmdalss, tkfkaskan1, gajh\}@snu.ac.kr, $^\ddagger$michael604@meta.com}

\author{
  \ifdefined\hpcacameraready
    \IEEEauthorblockN{\hpcaauthors{}}
      \IEEEauthorblockA{
        \hpcaaffiliation{} \\
        \hpcaemail{}
      }
  \else
    \IEEEauthorblockN{\normalsize{HPCA \hpcayear{} Submission
      \textbf{\#\hpcasubmissionnumber{}}} \\
      \IEEEauthorblockA{
        Confidential Draft \\
        Do NOT Distribute!!
      }
    }
  \fi 
}

\fancypagestyle{camerareadyfirstpage}{%
  \fancyhead{}
  
  \fancyhead[C]{
    \ifdefined\aeopen
    \parbox[][12mm][t]{13.5cm}{\hpcayear{} IEEE International Symposium on High-Performance Computer Architecture (HPCA)}    
    \else
      \ifdefined\aereviewed
      \parbox[][12mm][t]{13.5cm}{\hpcayear{} IEEE International Symposium on High-Performance Computer Architecture (HPCA)}
      \else
      \ifdefined\aereproduced
      \parbox[][12mm][t]{13.5cm}{\hpcayear{} IEEE International Symposium on High-Performance Computer Architecture (HPCA)}
      \else
      \parbox[][0mm][t]{13.5cm}{\hpcayear{} IEEE International Symposium on High-Performance Computer Architecture (HPCA)}
    \fi 
    \fi 
    \fi 
    \ifdefined\aeopen 
      \includegraphics[width=12mm,height=12mm]{ae-badges/open-research-objects.pdf}
    \fi 
    \ifdefined\aereviewed
      \includegraphics[width=12mm,height=12mm]{ae-badges/research-objects-reviewed.pdf}
    \fi 
    \ifdefined\aereproduced
      \includegraphics[width=12mm,height=12mm]{ae-badges/results-reproduced.pdf}
    \fi
  }
  \fancyfoot[C]{}
}
\begin{document} 
\maketitle

\ifdefined\hpcacameraready 
  \pagestyle{empty}
\else
  \thispagestyle{plain}
  \pagestyle{plain}
\fi

\newcommand{\hpcaheight}{0mm}
\ifdefined\eaopen
\renewcommand{\hpcaheight}{12mm}
\fi

\begingroup
  \renewcommand\thefootnote{}        
  \footnotetext{$^{*}$ Both authors contributed equally to the paper.}
  \addtocounter{footnote}{-1}       
\endgroup




\begin{abstract}

Modern HBM-based memory systems have evolved over generations while retaining cache line granularity accesses.
Preserving this fine granularity necessitated the introduction of bank groups and pseudo channels. 
These structures expand timing parameters and control overhead, significantly increasing memory controller scheduling complexity.
Large language models (LLMs) now dominate deep learning workloads, streaming contiguous data blocks ranging from several kilobytes to megabytes per operation.
%
In a conventional HBM-based memory system, these transfers are fragmented into hundreds of 32B cache line transactions.
This forces the memory controller to employ unnecessarily intricate scheduling, leading to growing inefficiency.

To address this problem, we propose \name.
\name accesses DRAM at row granularity and removes columns, bank groups, and pseudo channels from the memory interface.
This design simplifies memory scheduling, thereby requiring fewer pins per channel.
The freed pins are aggregated to form additional channels, increasing overall bandwidth by 12.5\% with minimal extra pins.
%
\name demonstrates how memory scheduling logic can be significantly simplified for representative LLM workloads, and presents an alternative approach for next-generation HBM-based memory systems achieving increased bandwidth with minimal hardware overhead.
%
%

\end{abstract}



\section{Introduction}
\label{sec:1_introduction}


High bandwidth memory (HBM) has emerged as a key component of high-performance computing systems~\cite{nvidia-2020-a100,nvidia-h100,nvidia-b200,isca-2023-tpuv4} driving the transformer-based artificial intelligence (AI) proliferation~\cite{neurips-2017-transformer}.
The high bandwidth of HBM is required to keep pace with the compute capabilities of GPUs~\cite{nvidia-b200, isca-2024-amd}, TPUs~\cite{isca-2023-tpuv4,google-2025-tpuv7}, and other AI accelerators~\cite{isca-2024-tcp, socc-2024-aws}, and satisfy the bandwidth-bound nature of the generation stages of the transformers~\cite{asplos-2024-attacc,asplos-2024-neupims,micro-2024-duplex}.
A single cube of HBM4~\cite{jedec-2025-hbm4} feeds 64 pseudo channels, each of which has 32 8\,Gbps data pins, constituting a total of 2\,TB/s bandwidth.

Saturating the immense HBM channel bandwidth requires efficient row-buffer utilization.
A DRAM bank is composed of a 2-dimensional array of DRAM cells and indexed by row and column addresses.
DRAM cell access latency is slow as it requires preceding precharge (PRE) and activation (ACT).
This latency is too high to saturate such a huge channel bandwidth.
%
%
To amortize those overheads, the bank prefetches an entire row into the \rowbuffer, which is orders of magnitude wider than the conventional DRAM access granularity.
This strategy is highly effective in saturating the DRAM channel bandwidth when there exists a large spatial locality in the memory access pattern (\eg, streaming access).
For example, when a series of memory reads (RDs) target the same \rowbuffer, the processor-side memory controller (MC) can issue back-to-back RDs to saturate the channel bandwidth without the bubbles induced by ACT/PRE.

Instead of issuing multiple consecutive read commands, we ask: why can't DRAM access granularity simply increase to match the row-level granularity in this scenario?
However, memory access patterns vary, and increasing the minimum DRAM access granularity can cause \emph{overfetch problem} by reading unnecessary data, degrading effective bandwidth~\cite{micro-2013-locality-aware}.
%
%
Thus, conventional DRAM access granularity is set to match the cache line size of a processor (\eg, 32\,B or 64\,B).

Such a cache-line-sized access granularity complicates the memory controller (MC) architecture.
To support column-level accesses (\ie, RDs/WRs), an MC must maintain bank states and timing parameters.
Moreover, it must operate many banks in parallel to hide the latency overhead of ACT and PRE commands~\cite{isca-2000-fr-fcfs, isca-2008-parbs, sc-2006-designspace}.
Further complexity arises from the need to determine when to issue PRE after ACT based on access patterns (\ie, the page policy~\cite{isca-2000-fr-fcfs,intel-2024-pagepolicy}).

The HBM hierarchy also becomes increasingly complex due to the fine-grained access granularity.
While HBM bandwidth has improved steadily~\cite{isscc-2014-hbm, isscc-2018-hbm2, isscc-2022-hbm3, jedec-2025-hbm4}, the bandwidth per bank has remained nearly unchanged.
As each bank operates at the DRAM core frequency and transfers data at a fixed access granularity, memory bandwidth can fundamentally be increased in only two ways: 1) by enlarging the access granularity or 2) by increasing the DRAM core frequency.
However, the former is constrained by cache line size, and physical limitations prevent significantly increasing the latter.

To overcome these limits, additional hierarchical structures, such as bank group and pseudo channel (PC), were introduced to boost bandwidth~\cite{jedec-2017-ddr4,jedec-2018-hbm2}.
Bank groups combine multiple banks and deliver data to the I/O at a higher frequency, allowing transfers from different bank groups to overlap while maintaining cache line granularity.
%
%
%
PCs increase the number of channels by reducing each channel's width, improving bandwidth without increasing both DRAM core frequency and access granularity.
Unlike previous generations, HBM4 doubles total bandwidth primarily by doubling I/Os (and thus PCs) without modifying the per-channel width.
While effective in raising bandwidth, these mechanisms add significant complexity to memory controller scheduling and timing.

\begin{figure}[!tb]
  \center
  \includegraphics[width=0.96\columnwidth]{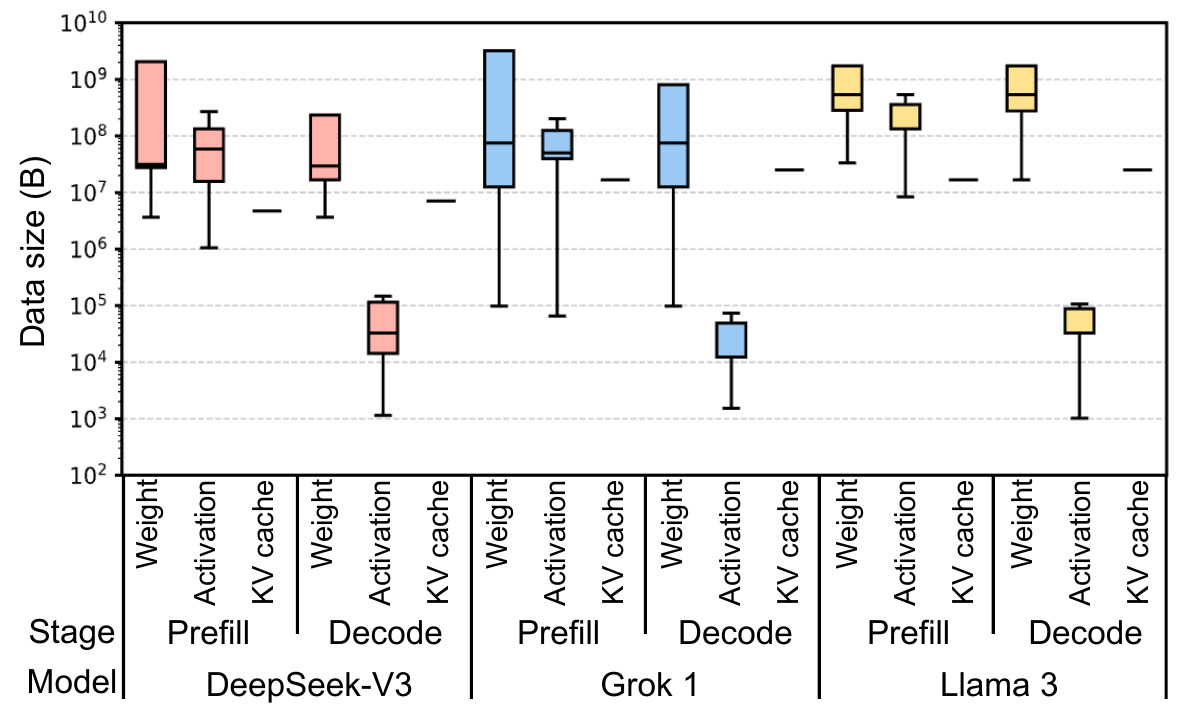}
  \vspace{-0.03in}
  \caption{Distribution of weight, activation, and KV cache size of DeepSeek-V3~\cite{arxiv-2024-deepseek-v3}, Grok 1~\cite{github-2024-grok1}, and Llama 3~\cite{arxiv-2024-llama3} in the \prefill and \decode stages.}
  \vspace{-0.1in}
  \label{fig:llm_access_gran}
\end{figure}

We challenge the need for such a conventional DRAM interface paradigm in the era of transformer-based Large Language Models (LLMs).
While traditional systems and data centers will persist, warehouse-scale systems running homogeneous applications of LLM inference (\eg, an AI factory~\cite{nvidia-ai-factory}) are becoming widespread.
LLMs mostly consist of simple general matrix-matrix or matrix-vector multiplication (GEMM/GEMV) and element-wise operations~\cite{asplos-2024-attacc}.
This is true even with state-of-the-art architectures, such as multi-head latent attention (MLA), grouped query attention (GQA), or mixture-of-experts (MoE)~\cite{arxiv-2024-deepseek-v2,arxiv-2023-llama2,github-2024-grok1}.
These models typically access tens of megabytes of data at once, far exceeding the size of conventional cache lines (Figure~\ref{fig:llm_access_gran}).
Unlike the workloads with irregular or strided patterns, LLM operations exhibit highly sequential memory access patterns.

By exploiting the memory access patterns of LLMs, we propose \name, a \textbf{Ro}w-granularity-access \textbf{Me}mory system designed to offer a simple and scalable memory system for LLM serving.
First, \name replaces a column-level access interface with a row-level one.
The sequence of DRAM commands required to access all data stored in a row is simplified into two commands: \rdrow and \wrrow.
Second, we propose a new bank architecture, virtual bank (\cba), to further simplify the interface by removing the bank group and PC in the MC-DRAM interface.
Given that memory accesses now operate at the row granularity, there is no longer a need to expose bank group and PC to the MC.
Accordingly, we design a single \cba to achieve maximum bandwidth without requiring any modification to the internal DRAM structure (\S\ref{sec:4_interface}).

Third, we introduce a command generator that decomposes each row-level command into a predefined sequence of conventional DRAM commands.
This enables the integration of additional HBM channels, improving overall memory bandwidth.
Finally, we demonstrate that the MC can be significantly simplified (\S\ref{subsec:5_2_our_mc}).
%
While conventional MCs rely on numerous data structures related to bank states and scheduling mechanisms to efficiently manage column-level commands, \name enables a significantly simpler and more scalable memory system design.

In a \name MC, five components are simplified: bank state, timing parameter, the number of bank finite-state machines (FSMs), request queue size, and scheduling algorithm.
%
The \name MC maintains only three bank states and fewer timing parameters, as the DRAM row-access command sequence is internally handled by the command generator that manages conventional timing parameters.
Two or fewer \cbas operate and up to three undergo refresh simultaneously; thus, just five bank FSMs need to be maintained.
This simplification enables a smaller request queue, as fewer in-flight requests need to be tracked.
Finally, as one \cba can provide maximum bandwidth, the scheduling algorithm is greatly simplified, focusing solely on interleaving across \cbas.

We evaluate \name using three representative LLMs: Grok 1~\cite{github-2024-grok1}, DeepSeek-V3~\cite{arxiv-2024-deepseek-v3}, and Llama 3~\cite{arxiv-2024-llama3}.
When serving LLM workloads, \name delivers higher performance than conventional HBM-based memory systems with significantly lower hardware overhead, while also providing modest gains in energy efficiency.
This demonstrates that, under the sequential and bulk access characteristics of LLM workloads, adopting row-level memory access does not degrade performance. 
While minor overheads may arise from overfetch and load imbalance, their impact is negligible.

The key contributions of this paper are as follows: 
\begin{itemize}[leftmargin=*,nolistsep]
\item We leverage the sequential and bulky memory access patterns of LLMs to propose a memory interface based on row-level access granularity.
\item As the cache-line-sized access granularity is no longer mandatory, we introduce a new bank architecture called \textit{virtual bank}, which eliminates the need for bank groups and pseudo channels.
\item A simplified memory controller optimized for the \name interface is designed to minimize control overhead associated with complex bank state tracking and scheduling, thereby reducing the area of the scheduling logic.
\item \name presents a method for expanding memory channels with minimal hardware overhead, thereby improving the performance of LLM workloads.
\end{itemize}

\section{Conventional memory systems}
\label{sec:2_background}


\subsection{Cache-Line-Sized DRAM Access Granularity}
\label{subsec:2_1_cacheline_access_gran}

Main-memory technologies such as DDR5~\cite{jedec-2024-ddr5} and HBM4, commonly integrated into modern CPUs and GPUs, are designed with access granularities that align with or are smaller than processor cache line sizes.
Specifically, HBM4 is optimized for 32B accesses, aligning with the cache line size of GPUs, while DDR5 supports 64\,B accesses, consistent with CPU cache line sizes.
Although DRAM rows are several kilobytes in size, these architectures enable fine-grained access at the column level, significantly smaller than the row size.

The adoption of cache-line-sized access granularity in main-memory systems serves two primary purposes.
First, aligning the access granularity with the processor’s cache line size minimizes data overfetch, thereby reducing unnecessary bandwidth usage and energy consumption by transferring only the data required by the program it executes.
Second, it enables flexibility in handling diverse memory access patterns.
This design effectively supports both sequential access patterns with high spatial locality, such as those LLMs, and random access patterns characterized by low spatial locality.

However, accessing memory at the cache line granularity introduces significant complexity in memory system design.
To achieve high performance, memory controllers (MCs) should implement sophisticated scheduling algorithms that account for a wide range of timing parameters and dynamic bank states.
Prior works have explored key components of this design space---including address mapping~\cite{micro-2000-interleaving}, page policies~\cite{micro-2011-minimalist,intel-2024-pagepolicy}, and scheduling policies~\cite{isca-2000-fr-fcfs,micro-2006-fair-queuing,micro-2007-stfm,isca-2008-parbs,micro-2010-TCM,hpca-2010-atlas}---which further contribute to the complexity of MC architecture.

\subsection{Bank Group \& Pseudo Channel}
\label{subsec:2_2_bg_pc}

\begin{figure}[!tb]
  \center
  \includegraphics[width=0.96\columnwidth]{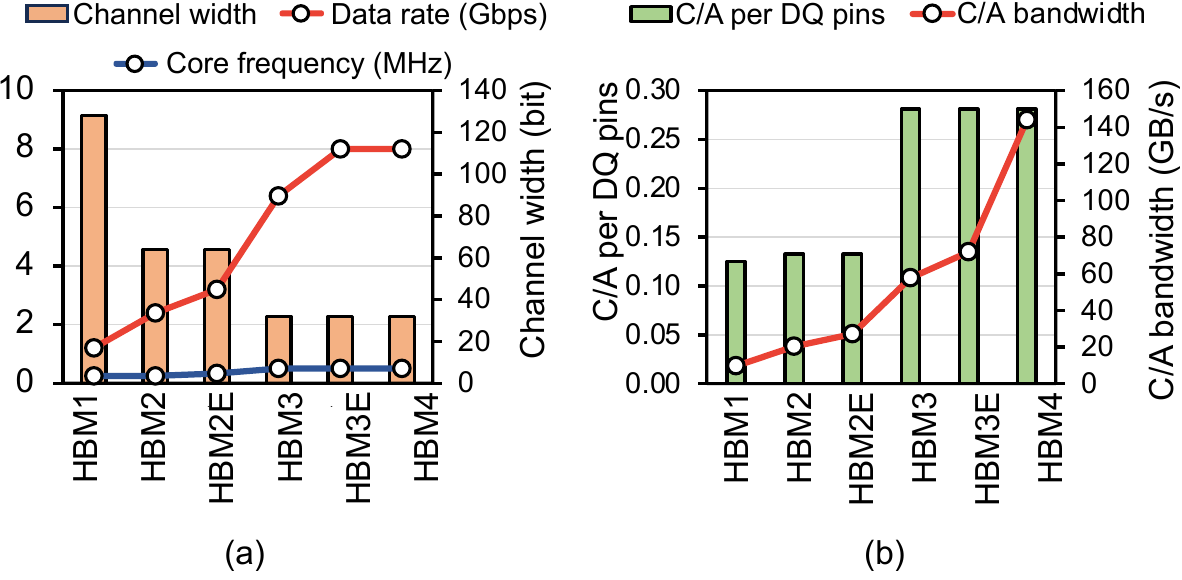}
  \vspace{-0.1in}
  \caption{(a) Trends in data rate, core frequency, and channel width, and (b) growth of C/A pin overhead across HBM generations.}
  \vspace{-0.1in}
  \label{fig:hbm_trend}
\end{figure}

Maintaining cache-line-sized access granularity adds complexity to the DRAM hierarchy, which in turn further increases the scheduling burden on the MC.
While DRAM bandwidth has steadily improved, maintaining fine-grained cache line access necessitated additional internal structures, specifically bank groups and pseudo channels (PCs).
As shown in Figure~\ref{fig:hbm_trend}(a), although the external data rate of DRAM devices has consistently increased, the DRAM core frequency has shown modest growth.
This limited scalability of core frequency is primarily due to the high energy and area overheads associated with its increase~\cite{micro-2017-fgdram}.
To meet the data rate demands under these constraints, a conventional approach has been to increase the amount of data fetched internally in a single access.

However, the increase in data rate leads to a mismatch between the access granularity and cache line size, prompting the introduction of the bank group structure~\cite{jedec-2017-ddr4}.
Instead of doubling the amount of data fetched from a single bank (\agbank), bank groups enable bandwidth scaling by alternating data fetches from banks in different bank groups at intervals of \tccds\ (typically equal to \tccdl/2), while preserving the cache-line-sized access granularity (\agmc).
This access strategy is referred to as bank group interleaving.
Each bank continues to operate at the DRAM core frequency (defined by \tccdl) and fetches data at the cache line granularity.
Thus, this mechanism allows effective scaling of the DRAM data rate without increasing \agbank or \agmc.
Terminologies are summarized in Table~\ref{tbl:symbols_background}.

Despite the introduction of the bank group structure, the demand for even higher external bandwidth has persisted, 
which led to the evolution of HBM toward narrower and more channels.
Figure~\ref{fig:hbm_trend}(a) illustrates this trend, showing a decrease in channel width and a corresponding increase in the number of channels across successive HBM generations~\cite{jedec-2013-hbm,jedec-2018-hbm2,jedec-2022-hbm3}.
In particular, each new generation of HBM has halved the channel width while doubling the number of channels.
As the data rate increases, the bandwidth per channel remains constant even with the narrower channel.
Notably, HBM4 scales the bandwidth by doubling the number of channels---therefore doubling the external I/O---without altering the channel width~\cite{jedec-2025-hbm4}.
This approach enables bandwidth scaling by populating more channels while maintaining per-channel bandwidth and preserving \agbank and \agmc.

\begin{table}[!tb]
    \centering
    \caption{Symbols and terminologies}
    \vspace{-0.05in}
    \label{tbl:symbols_background}
    \Scale[0.99]{
    \begin{tabular}{p{0.2\columnwidth} p{0.7\columnwidth}} 
        \Xhline{3\arrayrulewidth}
        \textbf{Symbol} & \textbf{Description} \\ 
        \Xhline{1.5\arrayrulewidth}
        \agbank   & Access granularity of a bank. \\
        \agmc     & Access granularity of a memory controller. \\ 
        \Xhline{0.5\arrayrulewidth}
        \Xhline{3\arrayrulewidth}
    \end{tabular}
    }
    \vspace{-0.1in}
\end{table}

However, these additional hierarchies exacerbate the scheduling complexity.
To fully utilize DRAM bandwidth, an MC must issue memory requests to different bank groups (\ie, bank group interleaving) and PCs.
This requires the MC to continuously track the state of all banks to identify those that are ready to accept new DRAM commands.
As a result, the MC must employ more sophisticated scheduling mechanisms to effectively leverage the complex DRAM hierarchy.

As the channel width narrows with each HBM generation, the overhead associated with command/address (C/A) pins increases.
HBM defines separate pins for row and column commands; for example, in HBM4, each 64-bit data channel requires 10 row command pins and 8 column command pins.
Moreover, populating more PCs proportionally increases independent C/A pins, raising the C/A-to-DQ pin ratio (see Figure~\ref{fig:hbm_trend}(b)).
From HBM1 and HBM2/2E to HBM3/3E and HBM4, this ratio has nearly doubled.
Further, the bandwidth requirements of C/A pins have steadily increased across generations, contributing to the rising overhead of the C/A interface.
Adopting these same techniques for future HBM generations with higher pin rates and bandwidths may be unsustainable.

\subsection{HBM Architecture}
\label{subsec:2_3_hbm}

\begin{figure}[!tb]
  \center
  \includegraphics[width=0.94\columnwidth]{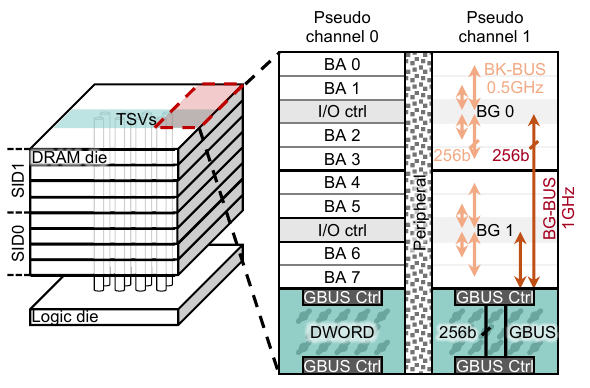}
  \vspace{-0.1in}
  \caption{Overview of HBM architecture and internal organization.}
  \vspace{-0.1in}
  \label{fig:hbm_architecture}
\end{figure}

HBM stacks multiple DRAM dies with a logic die at the bottom, which are connected by through silicon vias (TSVs), as shown in Figure~\ref{fig:hbm_architecture}.
%
Each HBM device is composed of multiple channels---up to 32 channels in the case of HBM4~\cite{jedec-2025-hbm4}---and forms a Stack ID (SID, equivalent to rank in conventional DRAM standards) for every four DRAM dies, supporting up to four SIDs per device. 
Each channel uses the SID to identify which group of DRAM dies it is accessing.
Each channel consists of two PCs, a design unique to HBM.
Two PCs in each channel share C/A pins but split the data pins evenly. 
The two PCs can operate independently, enabling concurrent data transfers and maximizing throughput.

Data transfer from individual banks within the DRAM dies to the logic die occurs as follows.
Each bank fetches 256 bits of data, corresponding to \agbank, and delivers it to the I/O control (ctrl) buffer via the bank data bus (BK-BUS).
Since all banks within a bank group share a single I/O control buffer, only one bank can occupy the BK-BUS at a time.
The data stored in the I/O ctrl buffer is then transferred over the bank group data bus (BG-BUS) to the global data bus (GBUS) controller and ultimately delivered to the logic die via the TSVs.
BK-BUS runs at the frequency of $1/\texttt{tCCDL}$ (\eg, 0.5 GHz), whereas BG-BUS runs at a faster frequency of $1/\texttt{tCCDS}$ (\eg, 1 GHz).
Therefore, a single bank group can utilize only half of the available bandwidth. To fully exploit the maximum bandwidth, data must be transmitted in a time-multiplexed manner across different bank groups.

\subsection{Conventional Memory Controller Architecture}
\label{subsec:2_4_conv_mc}

\begin{figure}[!tb]
  \center
  \includegraphics[width=0.99\columnwidth]{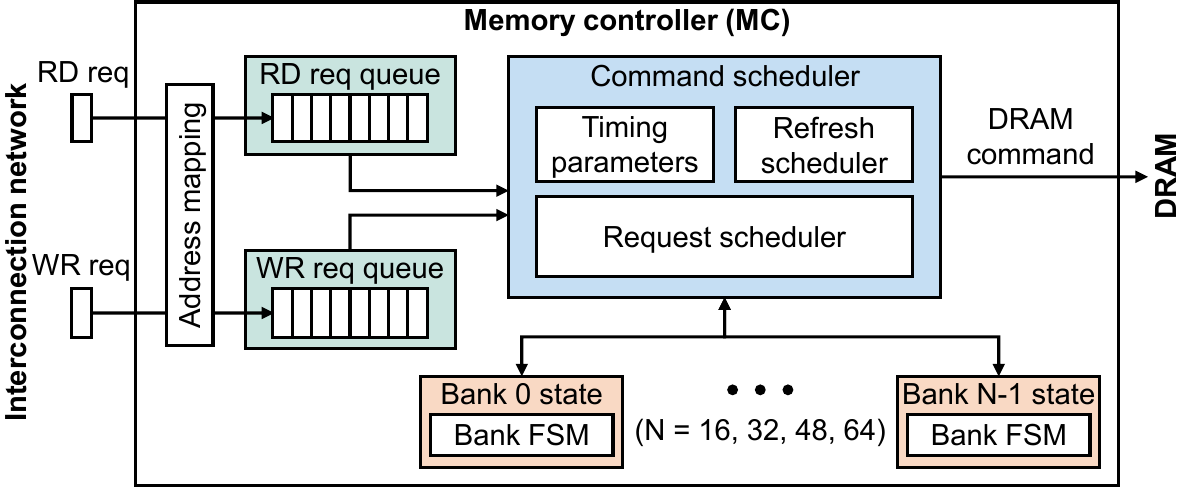}
  \vspace{-0.15in}
  \caption{Conventional memory controller architecture.}
  \vspace{-0.05in}
  \label{fig:MC}
\end{figure}

While implementation details vary, the high-level architecture of a generic MC is depicted in Figure~\ref{fig:MC}. 
MC generally includes four core components: address mapping, read/write request queue, per-bank state logic, and a command scheduler.
The address mapping unit translates the physical address of each read/write request received from the host into a corresponding DRAM address~\cite{isca-2024-dramscope,asplos-2025-marionette,sec-2016-drama,dac-2020-dramdig,dramsec-2025-sudoku,sec-2024-zenhammer} (\eg, PC and bank group) and inserts the translated request into the request queue.
Both the request queue and bank state logic are commonly implemented using content-addressable memory (CAM), allowing a one-cycle lookup to identify ready requests~\cite{isca-2012-sms}.
High bandwidth utilization requires a sufficiently large CAM to accommodate numerous in-flight requests.
As banks operate independently, per-bank state logic tracks the status of each bank.
The command scheduler is responsible for issuing memory and refresh commands by evaluating all bank states while adhering to DRAM timing constraints.
Each bank can be in one of seven states: Idle, Activating, Active, Precharging, Reading, Writing, and Refreshing.
%
The command scheduler must manage a wide range of timing parameters, which are summarized in Table~\ref{tab:hbm_timing}.
%
%

\renewcommand{\arraystretch}{1.2}
\begin{table}[!tb]
    \caption{Summary of HBM timing parameters}
    \vspace{-0.05in}
    \centering
    \label{tab:hbm_timing}
    \begin{tabular}{p{0.15\columnwidth}|p{0.75\columnwidth}}
        \Xhline{2\arrayrulewidth}
        \textbf{Parameter} & \textbf{Description} \\
        \Xhline{1.5\arrayrulewidth}
        tRCDRD    & ACT to RD delay in a same bank \\
        tRCDWR    & ACT to WR delay in a same bank \\
        tRAS      & ACT to PRE delay in a same bank \\
        tRP       & PRE to ACT delay in a same bank \\
        tCCDS(L/R)& RD/WR to RD/WR delay in diff BG (same BG/diff rank) \\
        tFAW      & Time window for 4 ACTs\\
        tRRDS(L)  & ACT to ACT delay in diff/same BG \\
        tWTRS(L)  & WR to RD delay in diff/same BG \\
        tRTW      & RD to WR delay in a same bank \\
        tWR       & WR to PRE delay in a same bank \\
        tRTP      & RD to PRE delay in a same bank \\
        \Xhline{2\arrayrulewidth}
    \end{tabular}
    \vspace{-0.1in}
\end{table}
\renewcommand{\arraystretch}{1.0}

Although the command scheduler performs various tasks, its responsibilities can be broadly categorized into refresh and request scheduling.
Refresh scheduler periodically issues refresh (REF) commands according to the \trefi interval, while optionally postponing or pooling REFs based on each bank's state~\cite{jedec-2025-hbm4}.

Request scheduler determines which request to schedule based on multiple criteria.
First, it exploits interleaving across banks, bank groups, and PCs. 
Bank interleaving helps hide ACT and PRE latencies by overlapping operations across independent banks.
Interleaving across bank groups and PCs further increases bandwidth utilization.
Second, the scheduler aims to exploit row buffer locality by issuing as many RDs/WRs as possible to an open row while obeying fairness; it pursues confining the overhead associated with ACT and PRE.
Third, it manages the page policy by determining the optimal time to precharge a row after activation, depending on memory access patterns.
This policy balances latency with row buffer hit rate and is typically implemented using open, close, or adaptive page policies~\cite{micro-2011-minimalist, isca-2014-rbd}.
Finally, to prevent starvation caused by the aggressive scheduling strategies, the scheduler incorporates Quality-of-Service (QoS~\cite{mkp-2004-ppin}) mechanisms that prioritize long-waiting requests, ensuring fairness across all memory transactions~\cite{micro-2007-stfm,isca-2008-parbs}.

\section{Access Pattern of Large Language Models}
\label{sec:3_motivation}

\begin{figure}[!tb]
  \center
  \includegraphics[width=0.97\columnwidth]{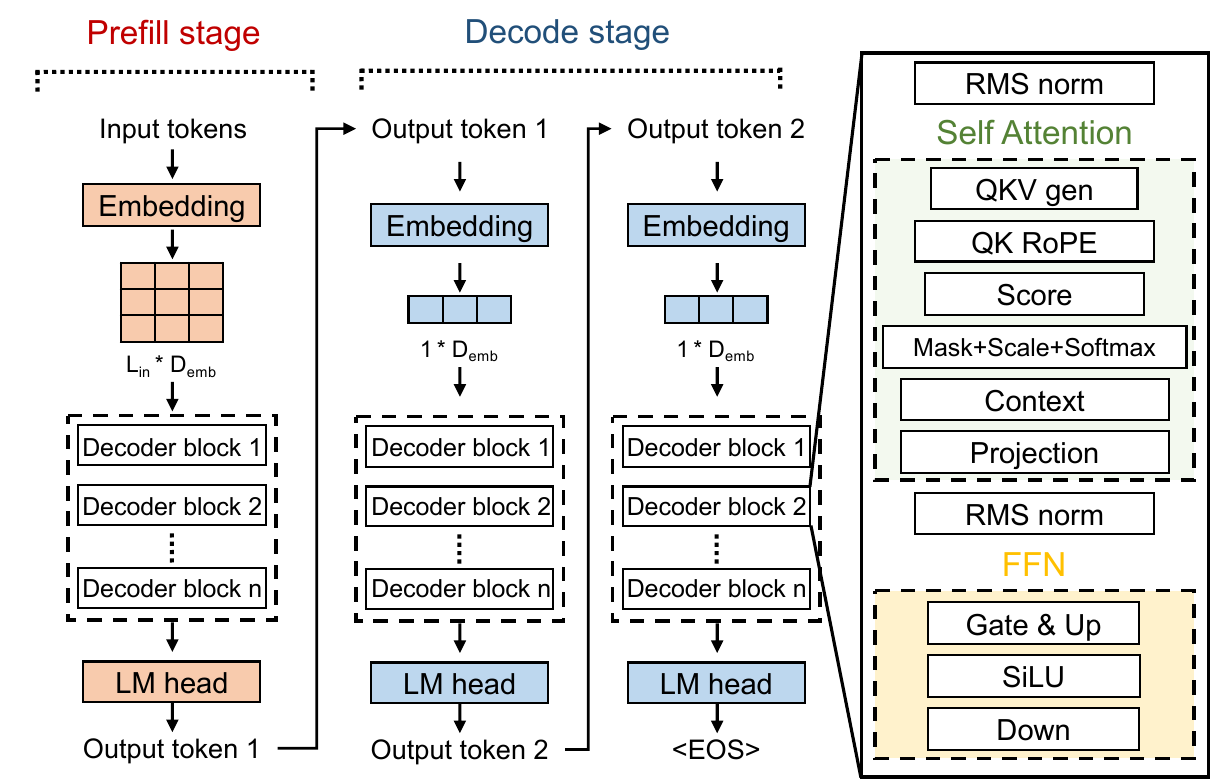}
  \vspace{-0.1in}
  \caption{Transformer-based LLM architecture.}
  \vspace{-0.15in}
  \label{fig:llm_arch}
\end{figure}

Widely adopted large language models (LLMs) are typically built upon the transformer decoder architecture (see Figure~\ref{fig:llm_arch}).
Throughout this paper, the term LLM refers specifically to a transformer-based LLM.
LLM inference can be broadly divided into two stages: \prefill and \decode.
In the \prefill stage, the model ingests all input tokens (e.g., words) in the request and generates the first output token.
In the \decode stage, it operates auto‑regressively, taking the output token from the previous step as input to produce the next output token.

Each stage consists of a token embedding layer, multiple decoder blocks, and a language model (LM) head. 
When an LLM processes an inference request containing a sequence of tokens, the embedding layer maps the input tokens into hidden vectors, which are then passed through the decoder blocks. 
Each decoder block takes the hidden vectors from the preceding decoder block and produces updated hidden vectors. 
Finally, the hidden vectors from the last decoder block are transformed into tokens by the LM head.
Here, we refer to the pre-trained model parameters (e.g., the weight of the fully-connected layer) as weight and the intermediate results of the operations and layers as activation.

In addition to the weights and activations, LLMs have a third primary data type: the KV‑cache.
Each decoder block is mainly composed of a self‑attention layer (attention) and a feed‑forward network (FFN).
The attention layer takes the hidden vector as input and produces the Query (Q), Key (K), and Value (V) matrices.
Because K and V store sequence context, the model requires K and V matrices for the entire sequence to generate each new token.
To avoid repeating the same computation at every generation step, the K and V matrices are stored in the KV‑cache.
Thus, the data used in LLM computation can be broadly categorized into weights, activations, and the KV‑cache.

During LLM execution, tens of megabytes of data typically need to be accessed sequentially at a time.
For all three LLMs in Figure~\ref{fig:llm_access_gran}, most weight and KV-cache accesses exceed several hundred kilobytes.
In Grok‑1, only one weight matrix is exceptionally small (24\,KB), but all other weight matrices exceed 12\,MB.
The KV‑cache also reaches several megabytes in the \decode stage; it grows even larger than in the \prefill stage because it must hold KV-cache for both the input and the already generated output tokens.
For activations, the \prefill stage processes all input tokens as a single batch, resulting in activation sizes reaching tens of megabytes.
In the \decode stage, however, only one token per sequence is processed, so the activation size is much smaller.
Nevertheless, given that modern LLM services often run with batch sizes in the hundreds~\cite{asplos-2024-attacc,asplos-2024-neupims,micro-2024-duplex,isca-2024-splitwise,osdi-2022-orca}, the activations can scale to a few megabytes, similar to the weights.

As GEMM and GEMV operations dominate LLM computations, these data are accessed with simple sequential memory access patterns.
However, current HBM-based memory systems are still designed for extremely fine‑grained 32\,B accesses, introducing unnecessary complexity relative to access characteristics of LLMs.
Therefore, we propose a highly simplified memory interface optimized for the sequential access pattern of LLMs and provide an in-depth analysis of its benefits.
We then present a co-optimization of DRAM and MC based on this interface, demonstrating a memory system for next‑generation AI accelerators that scales more effectively.

\begin{figure}[!tb]
  \center
  \includegraphics[width=0.96\columnwidth]{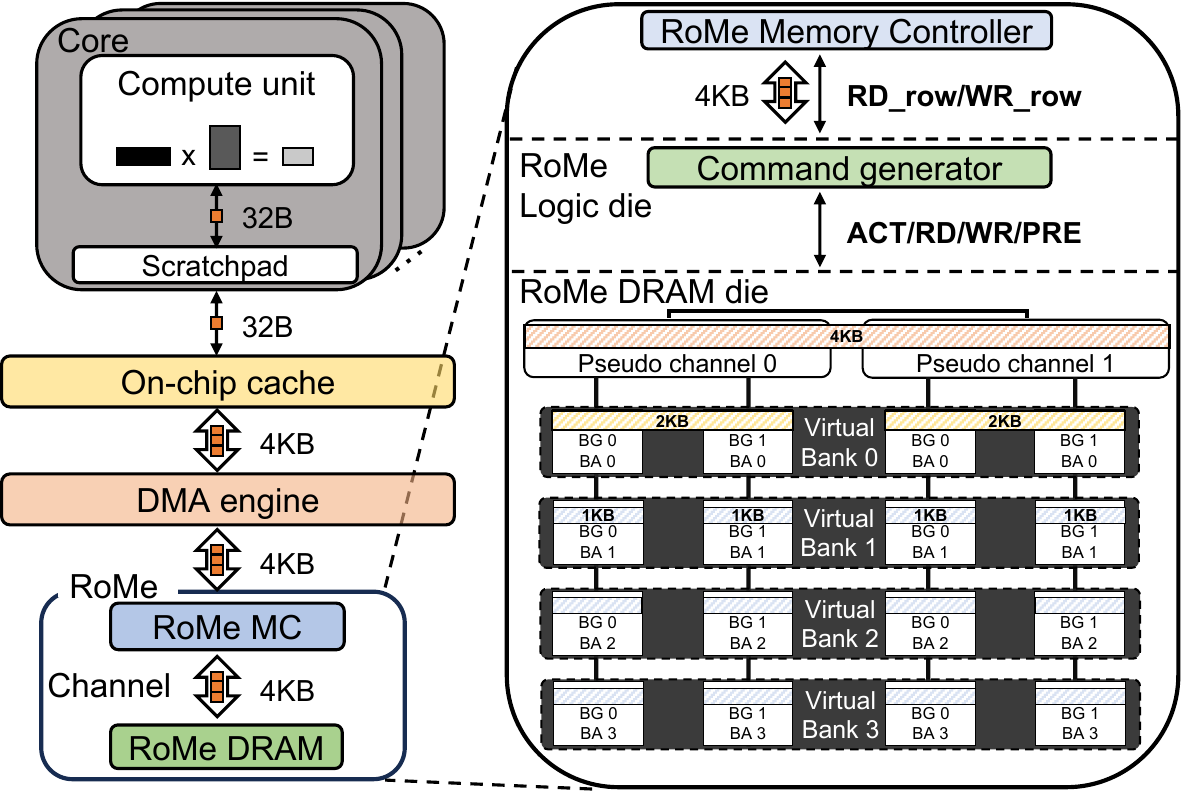}
  \vspace{-0.07in}
  
  \caption{An overview of a \name-based system.}
  \vspace{-0.1in}
  \label{fig:overview}
\end{figure}

\section{The \name interface}
\label{sec:4_interface}

%

\begin{figure*}[!tb]
  \center
  \includegraphics[width=0.9\textwidth]{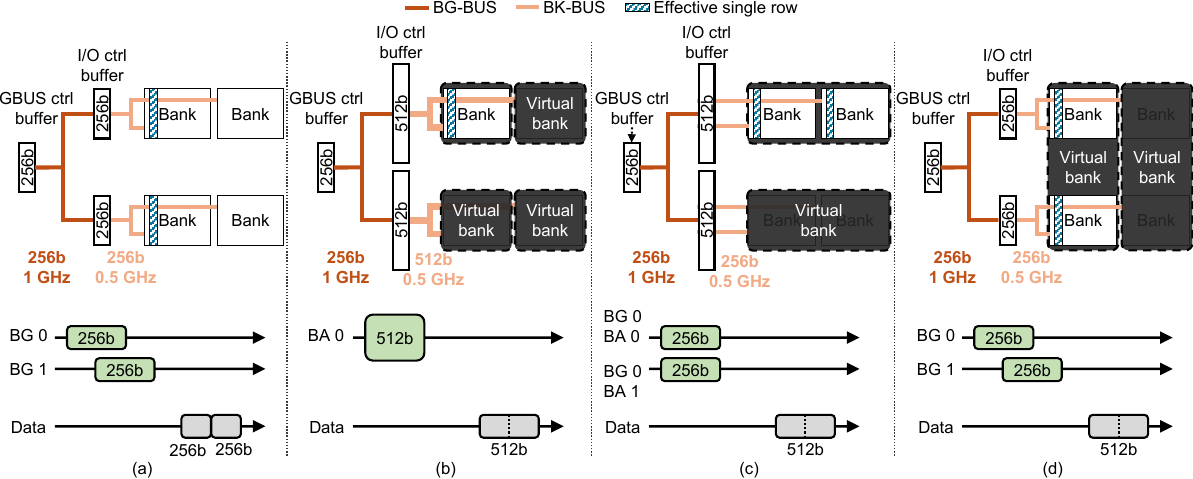}
  \vspace{-0.1in}
  \caption{Three design approaches to eliminate the bank group from the MC-DRAM interface. (a) Conventional bank group architecture. (b) A single bank serves as a \cba by doubling \agbank. (c) Two banks operate in tandem to form a \cba. (d) Two banks from different bank groups form a \cba and fetch data in an interleaved manner.}
  \vspace{-0.1in}
  \label{fig:bank_explore}
\end{figure*}

\begin{figure}[!tb]
  \center
  \includegraphics[width=0.96\columnwidth]{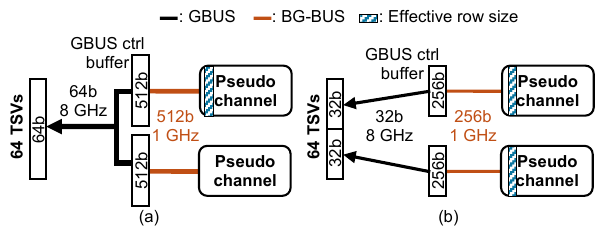}
  \vspace{-0.1in}
  \caption{Two design approaches to eliminate PC from the MC-DRAM interface. (a) A single PC operates as a full channel by fetching double the cache-line size. (b) Two PCs serve as a single channel and operate simultaneously, each maintaining data fetch size.}
  \vspace{-0.1in}
  \label{fig:pc_explore}
\end{figure}

\subsection{Memory Interface}
\label{subsec:4_1_memory_interface}

Exploiting the sequential and coarse-grained memory access patterns of LLM workloads, we propose a \textbf{Ro}w granularity access \textbf{Me}mory system, \name (Figure~\ref{fig:overview}).
For systems serving LLMs that sequentially access hundreds of megabytes of data at a time, the conventional cache-line-sized access granularity is excessively fine-grained.
\name replaces the traditional cache-line-level (column‑level) interface with a row‑level interface comprising \rdrow and \wrrow.
This increases \agmc from cache-line size to row size.
Because \agmc now corresponds to the row size, it is no longer necessary to align \agbank with the cache-line size.
Thus, we can further streamline the MC-DRAM interface by eliminating the bank group and PC from the interface that were originally introduced to scale bandwidth while retaining cache-line-sized \agbank.
With this significantly simplified interface, the \name MC no longer requires complex scheduling logic, leading to a much simpler architecture.

Moreover, we integrate command generators on the logic die to translate row-level commands into conventional DRAM commands.
This integration reduces the C/A pin count required per channel, allowing an HBM to add more channels with only a slightly increased pin budget, providing additional aggregate bandwidth.
%
Through both the simplified MC and the command generator, \name improves memory bandwidth with low hardware overhead, demonstrating decent scalability.
Their implications on area and energy are described in \S\ref{subsec:6_3_eval_energy}.

\name is designed to interoperate smoothly with modern AI accelerators.
LLM inference must continuously process massive weight and activation data, demanding not only high compute throughput, but also memory with both high bandwidth and high capacity.
Consequently, HBM has emerged as the standard memory for AI accelerators, and \name is accordingly designed to utilize the latest generation, HBM4.
Moreover, modern AI accelerators have adopted techniques that issue bulky memory accesses to efficiently fetch the enormous data required by AI workloads~\cite{isca-2023-tpuv4,nvidia-h100}.
In line with this trend, we assume a system where memory requests on the order of kilobytes are delivered to the MC.

\subsection{Virtual Bank}
\label{subsec:4_2_coarse_bank}

\name removes the concepts of bank group and PC, replacing them with a new hierarchy, a virtual bank (\cba).
The key idea behind \cba is to deliver the full available bandwidth from a single \cba, eliminating the need for complex MC-side scheduling that accounts for bank group or PC interleaving.
Because row granularity access no longer requires matching \agbank and \agmc to the cache-line size, traditional bank group and PC interfaces are no longer essential.
Accordingly, various design choices are possible for implementing \cba, and this work seeks to analyze the trade-offs associated with each.

There are three main design spaces for implementing a \cba that achieves maximum bandwidth from a single \cba.
First, as illustrated in Figure~\ref{fig:bank_explore}(b), a single bank can serve as a \cba by increasing its \agbank, thereby enabling it to deliver the maximum bandwidth.
While it maintains the same number of banks and effective row size, it requires doubling the bank's internal data path, the BK-BUS width, and the I/O ctrl buffer size, resulting in significant area overhead~\cite{micro-2017-fgdram}.
Second, as shown in Figure~\ref{fig:bank_explore}(c), a \cba consists of two banks within the same bank group.
By operating two banks in tandem, this approach fetches twice the amount of data, doubling the effective \agbank.
Although this does not change the internal data path, BK-BUS width, and I/O ctrl buffer, it effectively reduces the total number of banks by half and doubles the effective row size.
Finally, as shown in Figure~\ref{fig:bank_explore}(d), a \cba consists of two banks from different bank groups, accessed in a time-multiplexed manner.
This approach leverages the existing DRAM structure without modification while still enabling a single \cba to achieve maximum bandwidth.  
Similar to the second design, it reduces the number of banks by half and doubles the effective row size, but does so without requiring changes to the internal DRAM architecture.

There are two design approaches for eliminating the concept of PC.
Figure~\ref{fig:pc_explore}(a) illustrates a method where the amount of data fetched from each PC is doubled to enable a single PC to achieve maximum bandwidth.
However, this approach necessitates an increase in BG‑BUS width and the buffer size of the I/O ctrl buffer. 
Moreover, multiplexers are required between two GBUS on each PC side.
As a result, from the MC's perspective, the two PCs are controlled as a single channel, with the effective row size remaining 1KB while doubling the number of banks.
Figure~\ref{fig:pc_explore}(b) shows that both PCs operate simultaneously, similar to the legacy channel mode in HBM1/2~\cite{jedec-2013-hbm,jedec-2018-hbm2}.
This configuration doubles the bandwidth without requiring additional wiring and buffer, though the effective row size increases to 2\,KB.
In LLM workloads that involve fetching several megabytes of data, this increase in effective row size is not a significant issue.
Therefore, we eliminate the PC from the MC-DRAM interface by enabling the concurrent operation of two PCs (Figure~\ref{fig:pc_explore}(b)).

We conducted a comprehensive exploration of all combinations within the \cba design space, with the methodology and workloads detailed in \S\ref{subsec:6_1_methodology}.
%
%
Our experiments covered six total configurations, generated by combining the design points in Figure~\ref{fig:bank_explore}(b)/(c)/(d) with those in Figure~\ref{fig:pc_explore}(a)/(b).
Across all six configurations, the performance deviation relative to the baseline system remained within $3.6\%$.

However, the designs exhibit significant differences from the perspective of area overhead.
Using the configuration shown in Figure~\ref{fig:pc_explore}(a) requires doubling the width of the BG-BUS.
Similarly, the I/O ctrl buffer for Figure~\ref{fig:bank_explore}(b)/(c) must also be doubled, which in turn necessitates doubling the BK-BUS width for Figure~\ref{fig:bank_explore}(b).
%
When combined with the design in Figure~\ref{fig:bank_explore}(b)---where the BK-BUS and internal bank datalines are already doubled---the total dataline width becomes 4$\times$ that of traditional bank architecture, resulting in a substantial area overhead up to $77\%$~\cite{micro-2017-fgdram}.
Thus, we adopt the configuration in Figure~\ref{fig:bank_explore}(d) and Figure~\ref{fig:pc_explore}(b).

\begin{figure}[!tb]
  \center
  \includegraphics[width=0.96\columnwidth]{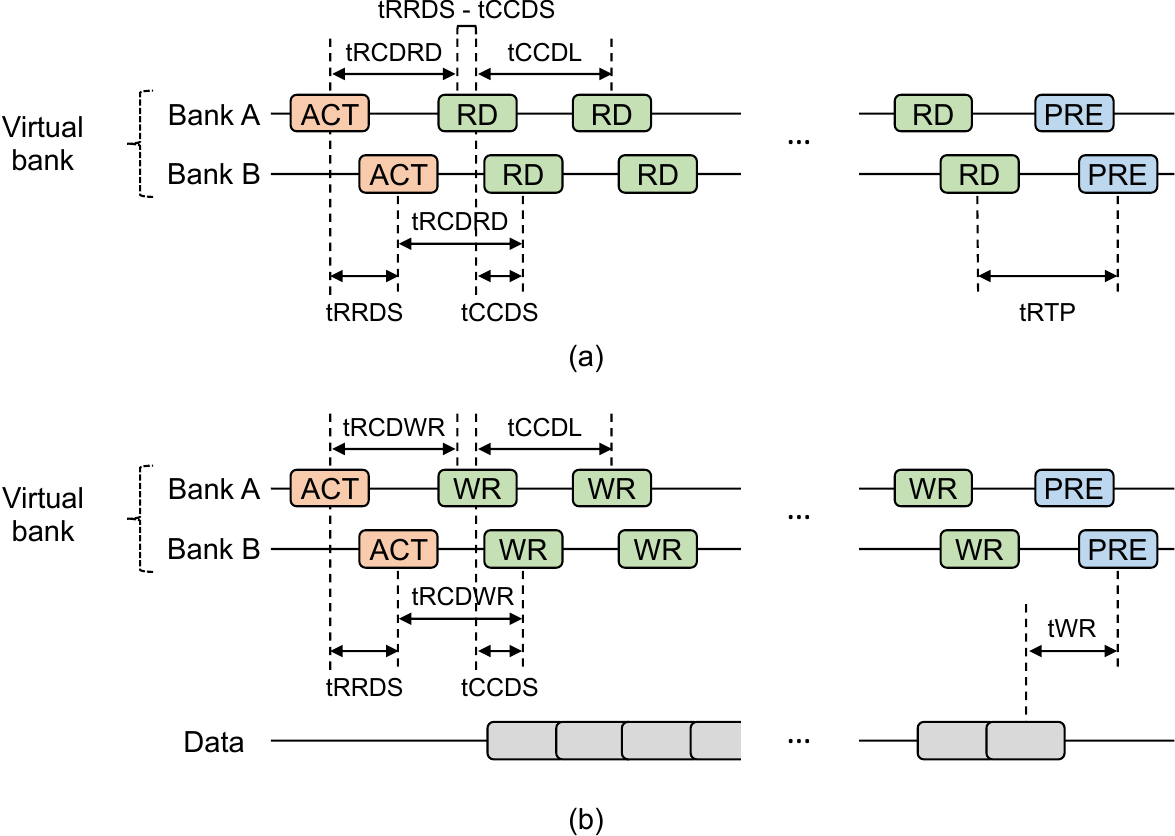}
  \vspace{-0.1in}
  \caption{Command sequence of (a) \rdrow and (b) \wrrow.}
  \vspace{-0.15in}
  \label{fig:command_sequence}
\end{figure}

\subsection{Command Generator}
\label{subsec:4_3_command_generator}

We add a command generator that accepts row-level commands and streams data from the \cba.
When the MC issues a \rdrow or \wrrow command, the command generator translates it into a fixed sequence of DRAM commands: one ACT, a series of RD or WR commands, and a PRE.
%
Unlike a conventional MC, our command generator does not issue commands dynamically based on bank states or timing constraints. 
Instead, it issues predetermined DRAM commands at fixed intervals upon receiving a row-level command, operating in a simplified and static manner.
Figure~\ref{fig:execution_flow} illustrates the detailed command sequences corresponding to \rdrow and \wrrow.
In \name, as in the legacy channel mode of HBM1/2, commands are sent to both PCs and data are also received from both simultaneously.
Because two PCs share the same C/A pins, we depict the command sequences for a single PC.

The command generator is designed to issue DRAM commands to two banks in a perfectly interleaved manner, ensuring that each RD/WR complies with \tccds (\eg, 1\,ns) between consecutive RD/WR to a different bank.
However, due to the \trrds constraint (\eg, 2\,ns), which must be satisfied between ACTs to different banks, maintaining this interleaving necessitates additional delay.
If both banks issue ACT followed by RD after \trcdrd, the RDs to different banks would align simultaneously rather than being interleaved. 
To resolve this, an intentional delay of \trrds $-$ \tccds is inserted before the ACT to the first bank (Figure~\ref{fig:command_sequence}).
This allows the RDs/WRs to the two banks to be issued at \tccds intervals.

The command generator can be placed in one of three locations: 1) MC, 2) logic die, or 3) DRAM die.
Placing the command generator in the MC has the benefit of minimizing modifications to the existing memory system.
However, this configuration limits the structural advantages that can be gained from a simplified memory interface, such as reducing C/A pins.
Integrating the command generator within the HBM stack helps reduce the C/A pin count between the MC and HBM.
When placed in the logic die, the command generator can reduce the C/A pin count between the MC and the logic die, though it does not reduce the number of TSVs between the logic and DRAM dies.
Placing it in the DRAM die can reduce TSV usage between the logic and DRAM dies, but it requires one command generator per channel for each DRAM die, increasing redundancy.

Given these trade-offs, we adopt a middle-ground design by placing the command generator in the logic die.
%
First, because the logic die of HBM4 is fabricated using a logic process (rather than a DRAM process)~\cite{samsung_base_die_process,hynix_base_die_process}, placing one command generator per channel incurs minimal area overhead (quantified in \S\ref{subsec:6_3_eval_energy}) while enabling effective reduction in C/A pin count.
Second, recent advances in die-stacking technologies, such as hybrid bonding~\cite{iedm-2020-lpddr-hybrid-bonding,isscc-2022-logic-hybrid-bonding}, help alleviate the cost associated with inter-die TSVs, making the logic-die placement a practical compromise.

\subsection{Command/Address Pins}
\label{subsec:4_4_cmdaddr}

Row granularity access enables a drastic reduction in the number of C/A pins between the MC and DRAM.
%
First, because separate RD and WR column C/A pins are no longer required, eight column C/A pins can be removed.
The mode register set (MRS), which is traditionally sent over a column command, is now transmitted via row C/A pins.
%
Out of the ten row C/A pins, up to four pins are used for the opcode, leaving pins for the address.
\name retains all four opcode pins but reduces the number of address pins.
Since \name does not require PC bits and each \cba includes two banks, one of the bank address bits is also unnecessary.
Excluding ACT and PRE, there are eight row commands.
Adding MRS, \rdrow, and \wrrow increases the total command count to eleven.
In a column-granularity interface, column C/A pins must support issuing RD and WR commands to both PCs every \tccds, and row C/A pins must support ACT commands every \trrds.

\begin{figure}[!tb]
  \center
  \includegraphics[width=0.96\columnwidth]{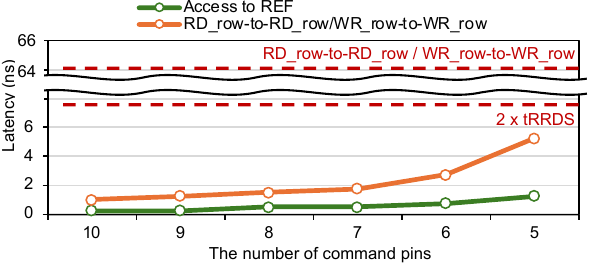}
  \vspace{-0.1in}
  \caption{Latency between \rdrow/\wrrow and REF across various numbers of C/A pins.}
  \vspace{-0.1in}
  \label{fig:ca_pin_overhead}
\end{figure}

However, with the row-level interface, the minimum interval between commands is longer.
The tightest timing occurs when a REF command is issued immediately after a \rdrow or \wrrow, requiring at least $2 \times t_{RRDS}$.
This is because one \trrds delay is needed between the ACT commands to the first and second banks, and another \trrds delay is needed before issuing the REF command to the second bank.
Figure~\ref{fig:ca_pin_overhead} shows the command issue latency as a function of the number of C/A pins.
Even with just five pins, commands can still be issued faster than $2 \times t_{RRDS}$. 
Therefore, by reducing the number of C/A pins to five, \name is able to eliminate 72\,\% of the C/A pins.

\subsection{Additional Channels}
\label{subsec:4_5_add_channel}

We utilize the freed C/A pin margin to introduce additional channels.
\name reduces the number of C/A pins from 18 to 5, saving 13 pins per channel.
A channel of HBM4 requires 120 pins~\cite{jedec-2025-hbm4}, whereas \name requires only 107 pins due to the 13-pin reduction.
Consequently, in a 32-channel configuration, 416 pins remain available, which allows the addition of four new channels with only 12 extra pins.
Through these additional channels, we aim to increase the memory bandwidth.

\name proposes to increase memory bandwidth by adding one additional channel per DRAM die.
As HBM generations have evolved, the number of channels per die has increased for channel expansion, necessitating a larger die area \cite{isscc-2014-hbm, isscc-2018-hbm2, isscc-2020-hbm2e, isscc-2022-hbm3, isscc-2024-hbm3e}.
Following this trend, \name also adopts a design expanding the number of channels per DRAM die from eight to nine.
As a result, \name-based HBM achieves approximately a 12.5\% increase in memory bandwidth merely with a small number of additional pins at the processor interface.
The area overhead is estimated in \S\ref{subsec:6_3_eval_energy}.

\section{Memory system under \name interface}
\label{sec:5_system}

\subsection{\name Memory Controller Architecture}
\label{subsec:5_2_our_mc}

As the memory interface is simplified, the MC can also be significantly simplified.
The MC now issues only three row-level commands (\rdrow, \wrrow, and REF), so the timing constraints among ACT, PRE, and RD/WR typical of conventional DRAM interfaces are eliminated.
Row-granularity operation also reduces the bank states in the bank FSM and timing parameters, and adopting \cba reduces the complexity of bank-state tracking logic.
Finally, the scheduler’s complexity for maximizing bandwidth is greatly reduced.

\begin{figure}[!tb]
  \center
  \includegraphics[width=0.9\columnwidth]{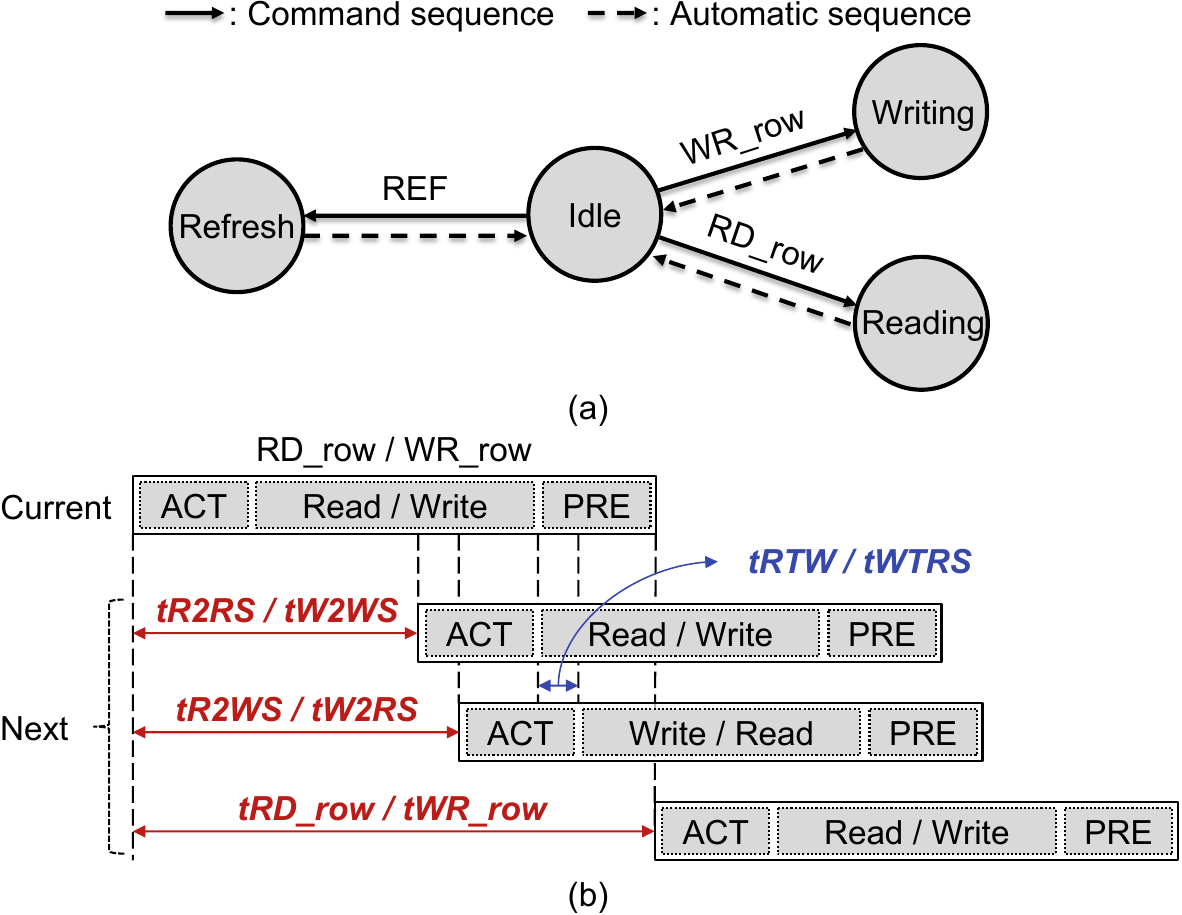}
  \vspace{-0.1in}
  \caption{(a) Bank state diagram and (b) timing parameters of \name MC.}
  \vspace{-0.1in}
  \label{fig:execution_flow}
\end{figure}

\noindent
\textbf{Bank states:}
Row-level access drastically simplifies the bank states related to data access.
Figure~\ref{fig:execution_flow}(a) illustrates the bank states of \name, which are Idle, Writing, Reading, and Refreshing.
Idle means the \cba is ready to accept a DRAM command immediately.
Refreshing indicates that a REF command is in progress.
Reading and Writing mean the bank is executing a \rdrow or \wrrow command, respectively. 
In conventional DRAM, after a Reading or Writing state, the bank returns to an Active state and can dynamically transition to additional reads or a precharge.
Under \name, however, DRAM is accessed only via \rdrow and \wrrow commands.
Upon completion, the bank automatically returns to Idle.
Thus, the Active, Activating, and Precharging states are no longer needed.

\begin{table}[!tb]
    \caption{Timing parameters of \name}
    \vspace{-0.05in}
    \centering
    \label{tab:reduced_timing}
        \begin{tabular}{p{0.17\columnwidth}|p{0.25\columnwidth}|p{0.35\columnwidth}}
        \Xhline{2\arrayrulewidth}
        \textbf{Name} & \textbf{Description} & \textbf{Destination} \\
        \Xhline{2\arrayrulewidth}
        \trtrs        & Different \cba       & \multirow{2}{*}{\rdrow to \rdrow} \\
        \trtrr        & Different SID \\
        \hline
        \trtws        & Different \cba       & \multirow{2}{*}{\rdrow to \wrrow} \\
        \trtwr        & Different SID \\
        \hline
        \twtrs        & Different \cba       & \multirow{2}{*}{\wrrow to \rdrow} \\
        \twtrr        & Different SID \\
        \hline
        \twtws        & Different \cba       & \multirow{2}{*}{\wrrow to \wrrow} \\
        \twtwr        & Different SID \\
        \hline
        \trdrow       & Same \cba            & \rdrow delay \\
        \hline
        \twrrow       & Same \cba            & \wrrow delay \\
        \Xhline{2\arrayrulewidth}
        \end{tabular}
        \vspace{-0.1in}
\end{table}

\noindent
\textbf{Timing parameters:}
The \name MC considers only a minimal set of timing constraints when performing memory accesses.
It issues only \rdrow, \wrrow, and REF commands, and each command returns its bank to the Idle state automatically.
Thus, it must track only the timing relationships between \rdrow and \wrrow commands. 
Rather than juggling the full set of row C/A and column C/A timing parameters in a conventional interface, the MC needs to manage only a few timing parameters. 
Table~\ref{tab:reduced_timing} lists the ten timing parameters used in \name, categorized by Read-to-Read, Read-to-Write, Write-to-Read, and Write-to-Write for same-bank, different-bank, and different-stack-ID cases.
Figure~\ref{fig:execution_flow}(b) illustrates when each timing parameter applies.
For \trtrs and \twtws, the next data transfer to the same row can begin immediately after the current one finishes.
For \trtws and \twtrs, the bus direction must switch, so an additional \trtw or \twtrss delay is incurred.
Accesses to different stack IDs (\trtrr, \trtwr, \twtwr, and \twtrr) incur a 1-2\,nCK longer delay than different-bank accesses~\cite{jedec-2025-hbm4}.
Finally, \trdrow and \twrrow simply chain within the same \cba, so the next operation can start as soon as the previous one completes.

\noindent
\textbf{The number of bank FSMs:}
Because \name drives at most two \cba at any given time, the MC needs only two bank FSM instances for scheduling.
Since two \cbas can saturate the bandwidth, the MC needs only to track the currently accessed \cba and the next \cba.
Nevertheless, due to the use of per-bank refresh, additional bank FSMs are implemented to track the status of banks being refreshed for a duration of tRFCpb divided by tREFIpb.
Memory requests are mapped to whichever bank FSM instance is free, and once a request completes, that FSM is deallocated.


\renewcommand{\arraystretch}{1.2}
\begin{table}[!tb]
    \caption{Simplified components of \name MC}
    \vspace{-0.05in}
    \centering
    \label{tab:simplifiedMC}
    \begin{tabular} {c|c|c}
        \Xhline{2\arrayrulewidth}
        \textbf{} & \textbf{Conventional MC} & \textbf{\name MC} \\
        \Xhline{2\arrayrulewidth}
        \# of timing params. & 15 & 10 \\
        \hline
        \# of bank FSMs & \# of total banks per PC & 5 \\
        \hline
        \# of bank states & 7 & 4 \\
        \hline
        Page policy & Open  & - \\
        \hline
        \multirow{3}{*}{Scheduling} & Row-buffer locality,      &  \multirow{3}{*}{\cba interleaving} \\
                                    & Bank group interleaving   &                    \\
                                    & PC interleaving           &                    \\
        \Xhline{2\arrayrulewidth}
    \end{tabular}
    \vspace{-0.1in}
\end{table}
\renewcommand{\arraystretch}{1.0}

\noindent
\textbf{Request queue size:}
The \name MC employing a highly simplified scheduler treats each 4KB access as a single request, enabling it to saturate DRAM bandwidth with a significantly smaller request queue.
In a cache-line access granularity, the ratio of \tccds to \trc exceeds 40$\times$, whereas with row granularity access, the ratio of \trtrs to \trdrow is less than 2$\times$.
If the queue size is too small, it cannot look far enough ahead to exploit bank-level parallelism; thus, a certain minimum size is still necessary.
%
%
HBM4 requires a queue depth of at least 45 entries, while \name achieves peak throughput with only two entries.
Thus, \name can saturate DRAM bandwidth with a depth of just two, allowing the MC to reduce the request queue size.

\noindent
\textbf{Command scheduling:}
The command scheduler delivers high performance and fairness with minimal complexity.
It first checks which \cba is active, then serves ready requests in oldest-first order.
As row-buffer locality is guaranteed by row granularity access, the scheduler needs only to avoid back-to-back commands to the same \cba to fully utilize bandwidth.
An age-based mechanism ensures that the oldest pending request is served next, improving tail latency and fairness.

Row-level access removes the need for any page-policy mechanism.
Conventional MCs dynamically switch between open, close, and adaptive page policies by monitoring row-buffer hits to adapt to varying access patterns~\cite{intel-2024-pagepolicy}.
In contrast, \name always precharges immediately after reading a row, inherently matching LLM's sequential access without requiring any additional policy logic.

\subsection{Refresh and Write Operations}
\label{subsec:5_3_ref_wr}
We optimize refresh behavior to suit the simplified interface with \cba.
For all-bank refresh (REFab), no bank in the target channel can operate during a refresh; thus, both the baseline and the \name MC behave the same.
By contrast, for per-bank refresh (REFpb), triggering a REF command on any single bank within a \cba blocks the entire \cba. 
Thus, it is important to minimize this overhead.
Instead of issuing a REFpb every \trefipb, MC issues one per-bank refresh every $2\times$\trefipb.
The command generator then sends two REFpb commands (one to each bank in the \cba) with an interval defined by the REFpb-to-REFpb timing interval, \trrefd.
This reduces the stall time per \cba from $2 \times \trfcpb$ (e.g., 2 × 280\,ns) to $\trfcpb + \trrefd$ (e.g., 280\,ns + 8\,ns).

Buffering multiple 4 KB write chunks in a write queue would require a large write buffer.
To avoid this, \name processes write requests immediately upon arrival, keeping the queue size small.
Since LLM workloads are heavily dominated by reads, the impact of immediate write handling is minimal.
Additionally, by issuing large 4 KB write requests atomically, \name reduces the frequency of read/write turnaround delays.

\section{Evaluation}
\label{sec:6_evaluation}

\subsection{Methodology}
\label{subsec:6_1_methodology}

\begin{table}[!tb]
    \caption{Timing parameters of HBM4 and \name}
    \vspace{-0.05in}
    \label{tbl:timing_param}
    \centering
    \resizebox{1\columnwidth}{!}{
        \begin{tabular}{l|c|c}
            \Xhline{2\arrayrulewidth}
                             & \textbf{HBM4} & \textbf{\name} \\ 
            \Xhline{2\arrayrulewidth}
            \begin{tabular}[c]{@{}c@{}}channels/cube \\ (PCs/cube)\end{tabular}     
                             & \begin{tabular}[c]{@{}c@{}}32 \\(64) \end{tabular}
                                             & 36            \\ \hline
            stacks           & 4             & 4             \\ \hline
            banks/channels   & 128           & 32            \\ \hline
            row size         & 1\,KB         & 4\,KB         \\ \hline
            data rate        & 8\,Gb/s       & 8\,Gb/s       \\ \hline
            bandwidth        & 2\,TB/s       & 2.25\,TB/s       \\ \hline
            timing parameter (ns) 
                & \begin{tabular}[c]{@{}c@{}}tRC=45, tRP=16, \\ tRAS=29, tCL=16, \\ tRCDRD=tRCDWR=16, \\ tWR=16, tFAW=12,
                                                             \\ tCCDL=2, tCCDS=1,\\ tCCDR=2, tRRD=2\end{tabular} 
                & \begin{tabular}[c]{@{}c@{}}
                                tR2RS/R=64/68 \\
                                tR2WS/R=69/73 \\
                                tW2RS/R=71/75 \\
                                tW2WS/R=64/68 \\
                                tRD\_row=95\\tWR\_row=115\end{tabular} 
                \\ \hline
            \agmc  & 32\,B & 4\,KB \\ 
            \Xhline{2\arrayrulewidth}
        \end{tabular}
    }
    \vspace{-0.1in}
\end{table}

\begin{figure*}[!tb]
  \center
  \includegraphics[width=0.94\textwidth]{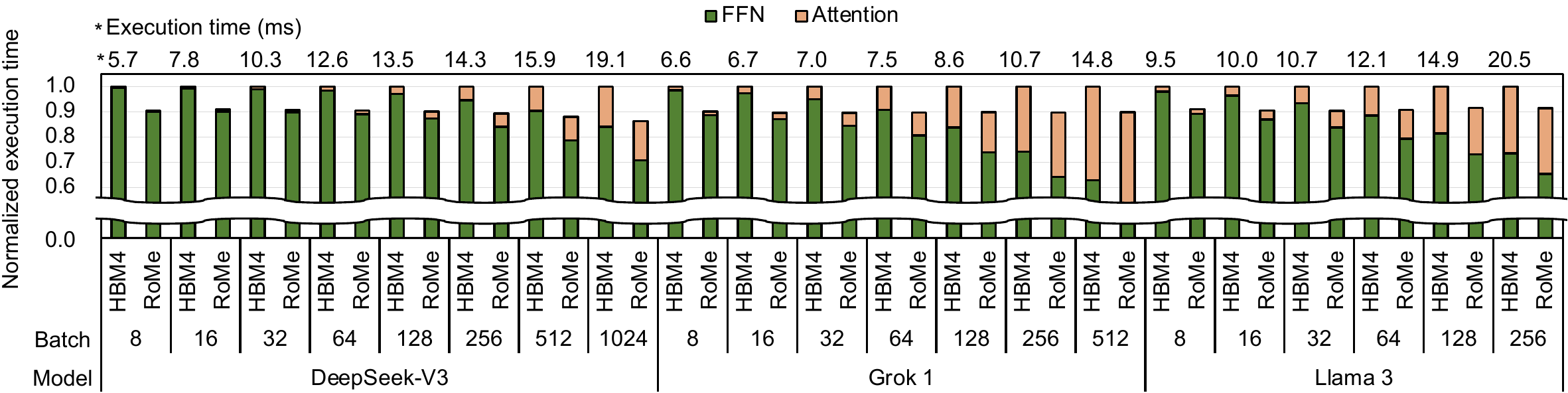}
  \caption{TPOT (time per output token) comparison between HBM4-based memory system and \name across various batch sizes for DeepSeek-V3, Grok 1, and Llama 3. The sequence length is 8K and the maximum batch size is constrained by memory capacity.}
  \vspace{-0.1in}
  \label{fig:tpot}
\end{figure*}

\noindent
\textbf{System:}
%
We first describe the configuration of a single accelerator and then extend the design to a multi-accelerator system representative of real LLM deployments.
Modern AI accelerators exhibit arithmetic intensities of 200–300\,Op/B for BF16 operations (\eg, 281 Op/B on B200~\cite{nvidia-b200}) and attach up to eight HBM cubes per device~\cite{nvidia-b200}.
Accordingly, we configure our target accelerator to sustain 280 Op/B and connect to eight HBM4 cubes.
Each HBM4 cube provides 32\,GB capacity with 8\,Gbps data rate and a 16‑Hi configuration~\cite{jedec-2025-hbm4}, yielding total 256\,GB memory system with 16 TB/s bandwidth.
To match our target arithmetic intensity, we scale BF16 throughput to 4480\,TFLOPS.
Because real‑world LLM deployments often span multiple devices to meet high capacity demands, we evaluate a system with eight accelerators operating in parallel, each providing 560 TFLOPS of BF16, 256\,GB of memory capacity, and 16\,TB/s of memory bandwidth.

\noindent
\textbf{Simulation:}
We model the AI accelerator equipped with the \name memory system, using LLMSimulator~\cite{micro-2024-duplex}.
It allows configuring both the accelerator and the memory subsystem, supports continuous batching, and integrates Ramulator 2.0~\cite{arxiv-2023-ramulator2} for cycle‑accurate DRAM simulation. 
We implement \name in Ramulator 2.0, configuring both the accelerator and \name to process 4\,KB requests.
From the simulator, we collect time per output token (TPOT) and DRAM energy.

To ensure fair comparison, we sweep address mappings for both the baseline and \name, selecting the configuration that maximizes bandwidth utilization.
We implement the MC for both systems using the FR‑FCFS scheduling policy~\cite{isca-2000-fr-fcfs}.
The baseline MC adopts an open‑page policy, while both systems employ per-bank refresh commands to improve bandwidth availability.
Table~\ref{tbl:timing_param} summarizes the timing parameters used in our experiments.
Because JEDEC has not finalized HBM4 timings, we adopt values from prior studies\cite{isca-2025-fbbank, micro-2017-fgdram}.

\noindent
\textbf{LLM:}
We evaluate three large‑scale LLMs: Grok 1~\cite{github-2024-grok1}, DeepSeek‑V3~\cite{arxiv-2024-deepseek-v3}, and Llama 3‑405B (Llama 3 hereafter)~\cite{arxiv-2024-llama3}.
DeepSeek-V3 uses Multi-head Latent Attention (MLA) and Mixture of Experts (MoE) together, Grok 1 adopts Grouped Query Attention (GQA) and MoE together, and Llama 3 adopts GQA but does not adopt MoE, instead using a fully-connected (FC) layer.
For MoE, DeepSeek‑V3 selects 8 of 256 experts per layer, while Grok 1 selects 2 of 8. 
All weights are stored in BF16.

During \prefill, we apply tensor parallelism (TP) across the eight accelerators.
During \decode, TP is applied to the attention layers with degrees of 1, 8, and 8 for DeepSeek‑V3, Grok 1, and Llama 3, respectively.
It is because the compressed KV cache of MLA favors data parallelism to avoid TP communication overhead\cite{arxiv-2025-mla}.
GQA runs with TP of 8, which our experiments and prior work have shown to be optimal\cite{arxiv-2024-llama3}.
For MoE, we use expert parallelism where each accelerator owns a distinct subset of experts, sending inputs to the target accelerator when a given expert is required and then receiving the output afterward.

\begin{figure}[!tb]
  \center
  \includegraphics[width=0.94\columnwidth]{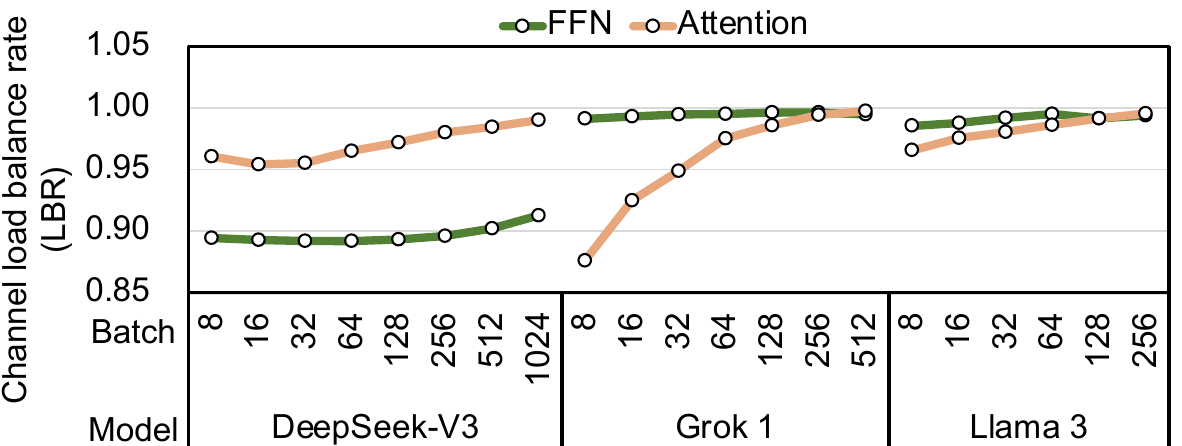}
  \vspace{-0.07in}
  \caption{Channel load balance ratio of \name across various batch sizes in DeepSeek-V3, Grok 1, and Llama 3 when the sequence length is 8K.} 
  \vspace{-0.1in}
  \label{fig:ch_util}
\end{figure}

\subsection{Performance Analysis of \name}
\label{subsec:6_2_eval_performance}

We measured the TPOT of the baseline (HBM4) and \name during the \decode stage with varying batch sizes when the sequence length is fixed at 8K.
As shown in Figure~\ref{fig:tpot}, \name reduces TPOT by $10.4\%$, $10.2\%$, and $9.0\%$ of HBM4 for DeepSeek-V3, Grok 1, and Llama 3, respectively.
This improvement is largely attributed to \name's 12.5\%
higher memory bandwidth from its increased number of channels.
However, the scaling does not fully align because several layers (\eg, FFN layers) are not memory-bound.

Because \name operates at a 4\,KB access granularity instead of the 32\,B, load imbalance across memory channels becomes a critical concern for effective bandwidth utilization.
Figure~\ref{fig:ch_util} shows channel load balance rate (\lbr) of \name for attention (\lbrattn) and FFN (\lbrffn) layers across various batch sizes.
\lbr quantifies how uniformly data is distributed across memory channels, with its values normalized to the HBM4 baseline, whose \lbr is nearly 1.
The value closer to 1 indicates a more uniform data distribution across memory channels, enabling \name to fully utilize its available bandwidth, while lower values reflect increasing imbalance.

\lbr differences across models primarily arise from their parallelization strategies and the relative contribution of weights and activations.
In the attention layers, the hidden dimensions are 7,168 (DeepSeek-V3), 6,144 (Grok 1), and 16,384 (Llama 3), which are proportional to weight sizes.
Given that data movement is dominated by weights at small batch sizes, DeepSeek-V3 adopts data parallelism, resulting in relatively high \lbrattn even with a small KV-cache size due to MLA.
%
In contrast, Grok~1 and Llama~3 employ TP and GQA, which reduces the data movement size of the weight per device, leading to lower \lbrattn at small batches.
However, Llama~3 still maintains high \lbrattn because its large hidden dimension size keeps the weight contribution significant even under TP.
As batch size increases, the KV-cache and activation footprints grow, improving \lbrattn across all three models.

\lbrffn is determined by their dimension size and architecture.
DeepSeek-V3, with a small intermediate dimension of 2,048, shows relatively low \lbrffn, while Grok~1 and Llama~3 have larger dimensions of 32,768 and 53,248, respectively.
For the FFN layers, DeepSeek-V3 and Grok~1 employ an MoE architecture, while Llama~3 uses a dense architecture.
In MoE layers, only a subset of experts (top-k) is activated, so \lbrffn improves only at large batch sizes where more experts are selected. 
\lbrffn improves once all experts begin to be selected, occurring around a batch of 64 in DeepSeek-V3 and a batch of 8 in Grok~1 in our experiments.

We omit results for the \prefill stage, as its performance remains unchanged under both memory systems due to its compute-bound nature.
This behavior stems from the characteristics of the \prefill stage.
Unlike the \decode stage, which typically processes a single input token at a time, the \prefill stage handles thousands of input tokens simultaneously.
As a result, the workload is dominated by GEMM operations.
Moreover, the large number of input tokens leads to significantly higher access volume to activations, weights, and KV-cache compared to the \decode stage.
Across all three evaluated LLMs, we observe that the performance difference in the \prefill stage remains within 0.1\%, confirming its insensitivity to the underlying memory system.

\subsection{Area \& Energy Overhead}
\label{subsec:6_3_eval_energy}
    
\begin{figure}[!tb]
  \center
  \includegraphics[width=0.97\columnwidth]{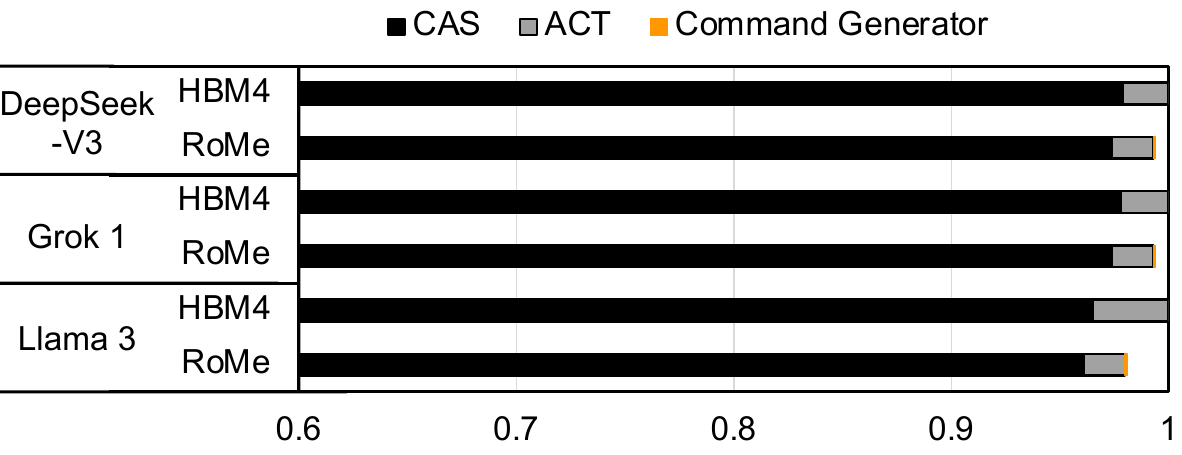}
  \vspace{-0.1in}
  \caption{Energy consumption of HBM4-based memory system and \name in DeepSeek-V3, Grok 1, and Llama 3 when the batch size is 256.}
  \vspace{-0.1in}
  \label{fig:energy_consumption}
\end{figure}


First, we calculated the area overhead incurred by the four additional channels based on HBM3E specifications~\cite{isscc-2024-hbm3e}.
The $\mu\text{bump}$ pitch was assumed to be $22~\mu\text{m}$~\cite{SKhynix-tsv-pitch}, which applies to both the DRAM and logic die. 
The number of $\mu\text{bumps}$ per channel was conservatively scaled by increasing it to four times the number required per channel~\cite{micro-2024-duplex,SKhynix-tsv-pitch}.
This configuration requires 48 additional $\mu\text{bump}$s for additional TSVs, corresponding to an area of approximately $0.14~\text{mm}^2$. 
Considering the edge margin, the DRAM die area increases by about $12\%$, and the logic die area grows proportionally, resulting in a total area overhead of only $0.10\%$.

We implemented the \name MC command scheduler and the command generator in Verilog, and utilized Synopsys Design Compiler with a $7$nm process technology~\cite{asap-7nm} to measure area and energy consumption.
%
%
Given that \name incorporates 36 legacy channels per cube, the total area overhead for the command generator amounts to $4268.8~\mu\text{m}^2$.
This represents negligible overhead, occupying 0.003\% of the logic die area.

For \name MC, we compared the area of the scheduling logic---including the command scheduler, bank FSM, and request queue---between \name and conventional MC.
The request queue depth was set to 64 entries for the conventional MC and 4 entries for \name MC, both evaluated under the FR-FCFS scheduling policy~\cite{isca-2000-fr-fcfs}.
Under these conditions, the command scheduling logic in \name MC occupies only $9.1\%$ of the area of a conventional MC, indicating that \name achieves a much simpler architecture.

Figure~\ref{fig:energy_consumption} shows the energy consumption of LLM workloads under HBM4 and \name.
We comprehensively calculated energy consumption by including the contributions of data movement within the HBM, command generator, and I/O interface.
The underlying energy model for HBM4 is adopted from \cite{isca-2025-fbbank}.
%
Compared to HBM4, \name reduces energy consumption by $1.9\%$, $0.7\%$, and $0.7\%$ for the three evaluated LLMs, respectively.
This improvement is primarily attributed to the decreased number of ACTs and the reduced energy consumption within the interposer.
Specifically, the ACT energy consumption is reduced to 55.5\%, 86.0\%, and 84.4\%, respectively.
Because \name accesses DRAM via \rdrow/\wrrow, it requires only the minimal number of ACTs regardless of the amount of data accessed, thereby minimizing energy consumption from ACTs.
Furthermore, interposer energy is reduced because \name MC issues a single \rdrow or \wrrow instead of 32 RDs or WRs.
Although overfetch may slightly increase the number of RDs and WRs, the overall overhead remains marginal.
Notably, the energy consumed by the command generator is negligible, contributing on average $0.06\%$ relative to the total energy consumption.
\emph{Overall, these results indicate that \name achieves slight improvements in energy efficiency while providing noticeable performance gains.}

\section{Discussion}
\label{sec:7_discussion}

\noindent
\textbf{Larger ECC codeword:}
\name may employ a larger ECC codeword by leveraging row access granularity (4\,KB).
%
HBM4 introduces two additional ECC pins per 32 DQ pins, building on the on-die ECC that has been available since HBM2E~\cite{jssc-2023-samsung-hbm3}.
Using a larger ECC codeword can reduce parity-bit overhead while maintaining comparable error correction and detection capabilities~\cite{isca-2011-adaptive-granularity,sc-2012-mage}.
This approach expands the design space for ECC, enabling the use of either conventional ECC schemes or more robust low-overhead alternatives.

\noindent
\textbf{Hybrid architecture for fine-grained access:}
\name is optimized for LLMs with coarse access granularity, but performance may degrade under workloads with frequent fine-grained accesses, such as sparse attention in gpt-oss~\cite{arxiv-2025-gptoss} and DeepSeek-V3.2~\cite{git-2025-sparseattn}. 
While gpt-oss maintains sequential access using sliding windows, DeepSeek Sparse Attention (DSA) selects top-2048 tokens from history, causing unpredictable and irregular access patterns when the sequence length exceeds 2048. 
This unpredictability can result in performance loss in \name due to overfetch.
To mitigate this, \name can adopt a heterogeneous system combining \name and conventional HBM4, assigning fine-grained requests to the latter. 
However, this may remain underutilized when processing fine-grained accesses, potentially leading to a reduction in overall bandwidth utilization.
Another approach is enhancing \name's memory controller and command generator to support selective column access via mask bits, though this introduces latency variation and added design complexity.

\noindent
\textbf{Processor-\name co-design:}
The effectiveness of \name critically depends on the processor's computational unit.
\name is optimized for row-granularity data movement, making it highly synergistic with processors that integrate a few large compute cores.
A key example is the Google TPU, which employs a limited number of brawny cores (\eg, 2 in TPUv7~\cite{google-2025-tpuv7}) with large shared on-chip buffers.
Such a design naturally benefits from \name's row access granularity by facilitating bulky and sequential data transfer.
Moreover, systems supporting explicit data management to and from the on-chip memory are preferred.
This programmability enables the processor to fully exploit \name's row access mechanism, maximizing throughput while efficiently controlling data movement.

\noindent
\textbf{\name for training:}
\name can be applied seamlessly to LLM training workloads.
Typically, LLM training operates on microbatches, where individual sequences are processed, each containing 8,192 tokens~\cite{sc-2021-microbatch,github-2025-trainpolicy}.
During the training process, multiple tokens are processed in parallel to maximize computational throughput.
This high degree of parallelism ensures that the memory access granularity is sufficiently large.
Consequently, \name is highly feasible in training scenarios.

\noindent
\textbf{Other types of DRAM:}
The \name approach can be applied to other DRAM types~\cite{micro-2024-smd} besides HBM.
First, the simplification of the MC architecture remains applicable, allowing a reduction in the area overhead associated with scheduling.
However, because only HBM incorporates a logic die, the placement of the command generator may differ, which could impose limitations on bandwidth expansion enabled by C/A pin reuse.
In addition, unlike HBM, conventional DRAM devices have a limited number of connectable pins.
Therefore, rather than aggregating the saved pins to construct additional channels as in \name, increasing the number of data pins could be a more effective approach for bandwidth expansion.

\section{Related Work}
\label{sec:8_related_work}

\noindent
\textbf{Coarse-grained access locality:}
Prior work has extensively documented the issue of memory access granularity arising from diverse application access patterns~\cite{sc-2006-designspace,sc-2012-mage,micro-2013-locality-aware,isca-2011-adaptive-granularity,isca-2012-dyanmic-granularity,iccd-2023-morpheus, asplos-2019-flatflash}.
Specifically,~\cite{sc-2012-mage,isca-2011-adaptive-granularity,isca-2012-dyanmic-granularity,micro-2013-locality-aware} identified problems regarding error correction, row buffer conflicts, and unnecessary data overfetch in systems that support a static granularity. 
They proposed memory systems that support adaptive memory access granularity for different applications. 
However, these designs do not take the memory access patterns specific to LLMs into account.
Further, because they keep the conventional memory interface unchanged, they still incur complex scheduling overhead.

\noindent
\textbf{Fine-grained access granularity:}
A body of work~\cite{asplos-2010-micro-pages,sc-2009-future-scaling,hpca-2017-architecting-energy-efficient,sc-2014-microbank,isca-2014-half-dram,islpled-2017-enabling-efficient-finegrained,hpca-2017-partial-row,isca-2010-rethinking,taco-2024-sectored-dram,micro-2016-improving-energh-efficiency,micro-2008-minirank, isca-2025-fbbank} has proposed energy-efficient fine-grained DRAM architectures.
\cite{micro-2008-minirank,sc-2009-future-scaling} enable DRAM chips within a DRAM module, which originally operate in lock-step, to function independently.
Fine-grained DRAM architectures~\cite{micro-2017-fgdram,hpca-2017-architecting-energy-efficient,sc-2014-microbank,isca-2014-half-dram,islpled-2017-enabling-efficient-finegrained,hpca-2017-partial-row, isca-2010-rethinking,micro-2016-improving-energh-efficiency} divide the DRAM bank into independently operating fine-grained memory arrays.
They primarily lower row activation energy by reducing the row size or saving access energy from data overfetch through finer access granularity, especially for applications with low spatial locality.
In contrast to prior work, \name focuses on the highly sequential memory access pattern of LLM inference, which both minimizes the overfetch overhead and fully exploits channel- and bank-level parallelism.
Through this approach, \name achieves a simplified memory interface, greater scalability, improved performance, and higher energy efficiency---even while increasing the row size and access granularity.

\section{Conclusion}
\label{sec:9_conclusion}

Motivated by the sequential and large memory traffic exhibited by large language models (LLMs), this paper proposes \name, a DRAM subsystem that adopts a row-granularity interface.
\name supports only row-level read and write commands, eliminating column, bank group, and pseudo channel layers required by conventional HBM and dispensing with cache line transfers entirely.
Freed from cache line granularity, we introduce the \cba organization, which efficiently supports row-level access.
We also present a command generator on the logic die that reduces the C/A pin requirements between the memory controller and HBM.
The freed pins are repurposed as additional channel interfaces, enabling efficient HBM channel expansion with minimal cost.
By simplifying the HBM hierarchy, \name lightens both the scheduling logic and hardware overhead of the memory controller.
Experiments on representative LLM workloads demonstrate that \name achieves higher performance and energy efficiency than HBM4 with minimal additional hardware overhead.
Any overfetch or load imbalance side effects introduced by row-granularity access remain negligible.

\section*{Acknowledgements}
We appreciate the input from Gunjun Lee at Seoul National University (SNU) regarding sparse attention.
This research was in part supported by Institute of Information \& communications Technology Planning \& Evaluation (IITP) grant funded by the Korea government (MSIT) [RS-2021-II211343, RS-2024-00456287, RS-2024-00402898, RS-2025-02304125], and
by the National Research Foundation of Korea (NRF) grant funded by MSIT [RS-2024-00405857]. 
The EDA tool was supported by the IC Design Education Center (IDEC), Korea.
This work was done when Michael Jaemin Kim was at SNU.
Jung Ho Ahn, the corresponding author, is with the Department of Intelligence and Information and the Interdisciplinary Program in Artificial Intelligence, SNU.

\balance
\bibliographystyle{IEEEtranS}
\bibliography{HPCA-2026/ref}

\begin{thebibliography}{10}
\providecommand{\url}[1]{#1}
\csname url@samestyle\endcsname
\providecommand{\newblock}{\relax}
\providecommand{\bibinfo}[2]{#2}
\providecommand{\BIBentrySTDinterwordspacing}{\spaceskip=0pt\relax}
\providecommand{\BIBentryALTinterwordstretchfactor}{4}
\providecommand{\BIBentryALTinterwordspacing}{\spaceskip=\fontdimen2\font plus
\BIBentryALTinterwordstretchfactor\fontdimen3\font minus \fontdimen4\font\relax}
\providecommand{\BIBforeignlanguage}[2]{{%
\expandafter\ifx\csname l@#1\endcsname\relax
\typeout{** WARNING: IEEEtranS.bst: No hyphenation pattern has been}%
\typeout{** loaded for the language `#1'. Using the pattern for}%
\typeout{** the default language instead.}%
\else
\language=\csname l@#1\endcsname
\fi
#2}}
\providecommand{\BIBdecl}{\relax}
\BIBdecl

\bibitem{asplos-2019-flatflash}
A.~Abulila, V.~S. Mailthody, Z.~Qureshi, J.~Huang, N.~S. Kim, J.~Xiong, and W.-m. Hwu, ``{FlatFlash: Exploiting the Byte-Accessibility of SSDs within a Unified Memory-Storage Hierarchy},'' in \emph{ASPLOS}, 2019.

\bibitem{isca-2025-fbbank}
V.~Adhinarayanan, B.~M. Beckmann, W.~Li, M.~Seyedzadeh, S.~Blagodurov, D.~Aguren, and H.~H. Lee, ``{Folded Banks: 3D-Stacked HBM Design for Fine-Grained Random-Access Bandwidth},'' in \emph{ISCA}, 2025.

\bibitem{sc-2006-designspace}
J.~Ahn, M.~Erez, and W.~J. Dally, ``{The Design Space of Data-Parallel Memory Systems},'' in \emph{SC}, 2006.

\bibitem{sc-2009-future-scaling}
J.~Ahn, N.~P. Jouppi, C.~Kozyrakis, J.~Leverich, and R.~S. Schreiber, ``{Future Scaling of Processor-Memory Interfaces},'' in \emph{SC}, 2009.

\bibitem{isca-2012-sms}
R.~Ausavarungnirun, K.~K.-W. Chang, L.~Subramanian, G.~H. Loh, and O.~Mutlu, ``{Staged Memory Scheduling: Achieving High Performance and Scalability in Heterogeneous Systems},'' in \emph{ISCA}, 2012.

\bibitem{asplos-2025-marionette}
S.~Baek, M.~Wi, S.~Park, H.~Nam, M.~J. Kim, N.~S. Kim, and J.~Ahn, ``{Marionette: A RowHammer Attack via Row Coupling},'' in \emph{{ASPLOS}}, 2025.

\bibitem{hpca-2017-architecting-energy-efficient}
N.~Chatterjee, M.~O'Connor, D.~Lee, D.~R. Johnson, S.~W. Keckler, M.~Rhu, and W.~J. Dally, ``{Architecting an Energy-Efficient DRAM System for GPUs},'' in \emph{HPCA}, 2017.

\bibitem{isscc-2018-hbm2}
J.~H. Cho, J.~Kim, W.~Y. Lee, D.~U. Lee, T.~K. Kim, H.~B. Park, C.~Jeong, M.-J. Park, S.~G. Baek, S.~Choi, B.~K. Yoon, Y.~J. Choi, K.~Y. Lee, D.~Shim, J.~Oh, J.~Kim, and S.-H. Lee, ``{A 1.2V 64Gb 341GB/s HBM2 Stacked DRAM with Spiral Point-to-Point TSV Structure and Improved Bank Group Data Control},'' in \emph{2018 IEEE International Solid-State Circuits Conference (ISSCC)}, 2018.

\bibitem{asap-7nm}
\BIBentryALTinterwordspacing
L.~T. Clark, V.~Vashishtha, L.~Shifren, A.~Gujja, S.~Sinha, B.~Cline, C.~Ramamurthy, and G.~Yeric, ``{ASAP7: A 7-nm finFET Predictive Process Design Kit},'' \emph{Microelectronics Journal}, 2016. [Online]. Available: \url{https://www.sciencedirect.com/science/article/pii/S002626921630026X}
\BIBentrySTDinterwordspacing

\bibitem{mkp-2004-ppin}
W.~J. Dally and B.~P. Towles, \emph{{Principles and Practices of Interconnection Networks}}.\hskip 1em plus 0.5em minus 0.4em\relax Morgan Kaufmann Publishers Inc., 2004.

\bibitem{git-2025-sparseattn}
\BIBentryALTinterwordspacing
DeepSeek-AI, ``{DeepSeek-V3.2-Exp: Boosting Long-Context Efficiency with DeepSeek Sparse Attention},'' 2025. [Online]. Available: \url{https://github.com/deepseek-ai/DeepSeek-V3.2-Exp}
\BIBentrySTDinterwordspacing

\bibitem{arxiv-2024-deepseek-v3}
\BIBentryALTinterwordspacing
DeepSeek-AI, A.~Liu, B.~Feng, B.~Xue, B.~Wang, B.~Wu, C.~Lu, C.~Zhao, C.~Deng, C.~Zhang, C.~Ruan, D.~Dai, D.~Guo, D.~Yang, D.~Chen, D.~Ji, E.~Li, F.~Lin, F.~Dai, F.~Luo, G.~Hao, G.~Chen, G.~Li, H.~Zhang, H.~Bao, H.~Xu, H.~Wang, H.~Zhang, H.~Ding, H.~Xin, H.~Gao, H.~Li, H.~Qu, J.~Cai, J.~Liang, J.~Guo, J.~Ni, J.~Li, J.~Wang, J.~Chen, J.~Chen, J.~Yuan, J.~Qiu, J.~Li, J.~Song, K.~Dong, K.~Hu, K.~Gao, K.~Guan, K.~Huang, K.~Yu, L.~Wang, L.~Zhang, L.~Xu, L.~Xia, L.~Zhao, L.~Wang, L.~Zhang, M.~Li, M.~Wang, M.~Zhang, M.~Zhang, M.~Tang, M.~Li, N.~Tian, P.~Huang, P.~Wang, P.~Zhang, Q.~Wang, Q.~Zhu, Q.~Chen, Q.~Du, R.~Chen, R.~Jin, R.~Ge, R.~Zhang, R.~Pan, R.~Wang, R.~Xu, R.~Zhang, R.~Chen, S.~Li, S.~Lu, S.~Zhou, S.~Chen, S.~Wu, S.~Ye, S.~Ye, S.~Ma, S.~Wang, S.~Zhou, S.~Yu, S.~Zhou, S.~Pan, T.~Wang, T.~Yun, T.~Pei, T.~Sun, W.~Xiao, W.~Zeng, W.~Zhao, W.~An, W.~Liu, W.~Liang, W.~Gao, W.~Yu, W.~Zhang, X.~Li, X.~Jin, X.~Wang, X.~Bi, X.~Liu, X.~Wang, X.~Shen, X.~Chen, X.~Zhang, X.~Chen, X.~Nie, X.~Sun, X.~Wang, X.~Cheng,
  X.~Liu, X.~Xie, X.~Liu, X.~Yu, X.~Song, X.~Shan, X.~Zhou, X.~Yang, X.~Li, X.~Su, X.~Lin, Y.~Li, Y.~Wang, Y.~Wei, Y.~Zhu, Y.~Zhang, Y.~Xu, Y.~Xu, Y.~Huang, Y.~Li, Y.~Zhao, Y.~Sun, Y.~Li, Y.~Wang, Y.~Yu, Y.~Zheng, Y.~Zhang, Y.~Shi, Y.~Xiong, Y.~He, Y.~Tang, Y.~Piao, Y.~Wang, Y.~Tan, Y.~Ma, Y.~Liu, Y.~Guo, Y.~Wu, Y.~Ou, Y.~Zhu, Y.~Wang, Y.~Gong, Y.~Zou, Y.~He, Y.~Zha, Y.~Xiong, Y.~Ma, Y.~Yan, Y.~Luo, Y.~You, Y.~Liu, Y.~Zhou, Z.~Wu, Z.~Ren, Z.~Ren, Z.~Sha, Z.~Fu, Z.~Xu, Z.~Huang, Z.~Zhang, Z.~Xie, Z.~Zhang, Z.~Hao, Z.~Gou, Z.~Ma, Z.~Yan, Z.~Shao, Z.~Xu, Z.~Wu, Z.~Zhang, Z.~Li, Z.~Gu, Z.~Zhu, Z.~Liu, Z.~Li, Z.~Xie, Z.~Song, Z.~Gao, and Z.~Pan, ``{DeepSeek-V3 Technical Report},'' 2024. [Online]. Available: \url{https://arxiv.org/abs/2412.19437}
\BIBentrySTDinterwordspacing

\bibitem{arxiv-2024-llama3}
A.~Dubey, A.~Jauhri, A.~Pandey, A.~Kadian, A.~Al-Dahle, A.~Letman, A.~Mathur, A.~Schelten, A.~Yang, A.~Fan, A.~Goyal, A.~Hartshorn, A.~Yang, A.~Mitra, A.~Sravankumar, A.~Korenev, A.~Hinsvark, A.~Rao, A.~Zhang, A.~Rodriguez, A.~Gregerson, A.~Spataru, B.~Roziere, B.~Biron, B.~Tang, B.~Chern, C.~Caucheteux, C.~Nayak, C.~Bi, C.~Marra, C.~McConnell, C.~Keller, C.~Touret, C.~Wu, C.~Wong, C.~Ferrer, Canton, C.~Nikolaidis, D.~Allonsius, D.~Song, D.~Pintz, D.~Livshits, D.~Esiobu, D.~Choudhary, D.~Mahajan, D.~Garcia-Olano, D.~Perino, D.~Hupkes, E.~Lakomkin, E.~AlBadawy, E.~Lobanova, E.~Dinan, E.~Smith, Michael, F.~Radenovic, F.~Zhang, G.~Synnaeve, G.~Lee, G.~Anderson, Lewis, G.~Nail, G.~Mialon, G.~Pang, G.~Cucurell, H.~Nguyen, H.~Korevaar, H.~Xu, H.~Touvron, I.~Zarov, I.~Ibarra, Arrieta, I.~Kloumann, I.~Misra, I.~Evtimov, J.~Copet, J.~Lee, J.~Geffert, J.~Vranes, J.~Park, J.~Mahadeokar, J.~Shah, J.~v.~d. Linde, J.~Billock, J.~Hong, J.~Lee, J.~Fu, J.~Chi, J.~Huang, J.~Liu, J.~Wang, J.~Yu, J.~Bitton, J.~Spisak, J.~Park,
  J.~Rocca, J.~Johnstun, J.~Saxe, J.~Jia, K.~Alwala, Vasuden, K.~Upasani, K.~Plawiak, K.~Li, K.~Heafield, K.~Stone, K.~El-Arini, K.~Iyer, K.~Malik, K.~Chiu, K.~Bhalla, L.~Rantala-Yeary, L.~v.~d. Maaten, L.~Chen, L.~Tan, L.~Jenkins, L.~Martin, L.~Madaan, L.~Malo, L.~Blecher, L.~Landzaat, L.~Oliveira, de, M.~Muzzi, M.~Pasupuleti, M.~Singh, M.~Paluri, M.~Kardas, M.~Oldham, M.~Rita, M.~Pavlova, M.~Kambadur, M.~Lewis, M.~Si, M.~Singh, Kumar, M.~Hassan, N.~Goyal, N.~Torabi, N.~Bashlykov, N.~Bogoychev, N.~Chatterji, O.~Duchenne, O.~Çelebi, P.~Alrassy, P.~Zhang, P.~Li, P.~Vasic, P.~Weng, P.~Bhargava, P.~Dubal, P.~Krishnan, P.~Koura, Singh, P.~Xu, Q.~He, Q.~Dong, R.~Srinivasan, R.~Ganapathy, R.~Calderer, R.~Cabral, Silveira, R.~Stojnic, R.~Raileanu, R.~Girdhar, R.~Patel, R.~Sauvestre, R.~Polidoro, R.~Sumbaly, R.~Taylor, R.~Silva, R.~Hou, R.~Wang, S.~Hosseini, S.~Chennabasappa, S.~Singh, S.~Bell, S.~Kim, Sonia, S.~Edunov, S.~Nie, S.~Narang, S.~Raparthy, S.~Shen, S.~Wan, S.~Bhosale, S.~Zhang, S.~Vandenhende, S.~Batra,
  S.~Whitman, S.~Sootla, S.~Collot, S.~Gururangan, S.~Borodinsky, T.~Herman, T.~Fowler, T.~Sheasha, T.~Georgiou, T.~Scialom, T.~Speckbacher, T.~Mihaylov, T.~Xiao, U.~Karn, V.~Goswami, V.~Gupta, V.~Ramanathan, V.~Kerkez, V.~Gonguet, V.~Do, V.~Vogeti, V.~Petrovic, W.~Chu, W.~Xiong, W.~Fu, W.~Meers, X.~Martinet, X.~Wang, X.~Tan, Ellen, X.~Xie, X.~Jia, X.~Wang, Y.~Goldschlag, Y.~Gaur, Y.~Babaei, Y.~Wen, Y.~Song, Y.~Zhang, Y.~Li, Y.~Mao, Z.~Coudert, Delpierre, Z.~Yan, Z.~Chen, Z.~Papakipos, A.~Singh, A.~Grattafiori, A.~Jain, A.~Kelsey, A.~Shajnfeld, A.~Gangidi, A.~Victoria, A.~Goldstand, A.~Menon, A.~Sharma, A.~Boesenberg, A.~Vaughan, A.~Baevski, A.~Feinstein, A.~Kallet, A.~Sangani, A.~Yunus, A.~Lupu, A.~Alvarado, A.~Caples, A.~Gu, A.~Ho, A.~Poulton, A.~Ryan, A.~Ramchandani, A.~Franco, A.~Saraf, A.~Chowdhury, A.~Gabriel, A.~Bharambe, A.~Eisenman, A.~Yazdan, B.~James, B.~Maurer, B.~Leonhardi, B.~Huang, B.~Loyd, B.~Paola, De, B.~Paranjape, B.~Liu, B.~Wu, B.~Ni, B.~Hancock, B.~Wasti, B.~Spence, B.~Stojkovic,
  B.~Gamido, B.~Montalvo, C.~Parker, C.~Burton, C.~Mejia, C.~Wang, C.~Kim, C.~Zhou, C.~Hu, C.-H. Chu, C.~Cai, C.~Tindal, C.~Feichtenhofer, D.~Civin, D.~Beaty, D.~Kreymer, D.~Li, D.~Wyatt, D.~Adkins, D.~Xu, D.~Testuggine, D.~David, D.~Parikh, D.~Liskovich, D.~Foss, D.~Wang, D.~Le, D.~Holland, E.~Dowling, E.~Jamil, E.~Montgomery, E.~Presani, E.~Hahn, E.~Wood, E.~Brinkman, E.~Arcaute, E.~Dunbar, E.~Smothers, F.~Sun, F.~Kreuk, F.~Tian, F.~Ozgenel, F.~Caggioni, F.~Guzmán, F.~Kanayet, F.~Seide, G.~Florez, Medina, G.~Schwarz, G.~Badeer, G.~Swee, G.~Halpern, G.~Thattai, G.~Herman, G.~Sizov, Guangyi, Zhang, G.~Lakshminarayanan, H.~Shojanazeri, H.~Zou, H.~Wang, H.~Zha, H.~Habeeb, H.~Rudolph, H.~Suk, H.~Aspegren, H.~Goldman, I.~Damlaj, I.~Molybog, I.~Tufanov, I.-E. Veliche, I.~Gat, J.~Weissman, J.~Geboski, J.~Kohli, J.~Asher, J.-B. Gaya, J.~Marcus, J.~Tang, J.~Chan, J.~Zhen, J.~Reizenstein, J.~Teboul, J.~Zhong, J.~Jin, J.~Yang, J.~Cummings, J.~Carvill, J.~Shepard, J.~McPhie, J.~Torres, J.~Ginsburg, J.~Wang, K.~Wu,
  K.~U, Hou, K.~Saxena, K.~Prasad, K.~Khandelwal, K.~Zand, K.~Matosich, K.~Veeraraghavan, K.~Michelena, K.~Li, K.~Huang, K.~Chawla, K.~Lakhotia, K.~Huang, L.~Chen, L.~Garg, L.~A, L.~Silva, L.~Bell, L.~Zhang, L.~Guo, L.~Yu, L.~Moshkovich, L.~Wehrstedt, M.~Khabsa, M.~Avalani, M.~Bhatt, M.~Tsimpoukelli, M.~Mankus, M.~Hasson, M.~Lennie, M.~Reso, M.~Groshev, M.~Naumov, M.~Lathi, M.~Keneally, M.~Seltzer, L., M.~Valko, M.~Restrepo, M.~Patel, M.~Vyatskov, M.~Samvelyan, M.~Clark, M.~Macey, M.~Wang, M.~Hermoso, Jubert, M.~Metanat, M.~Rastegari, M.~Bansal, N.~Santhanam, N.~Parks, N.~White, N.~Bawa, N.~Singhal, N.~Egebo, N.~Usunier, N.~Laptev, Pavlovich, N.~Dong, N.~Zhang, N.~Cheng, O.~Chernoguz, O.~Hart, O.~Salpekar, O.~Kalinli, P.~Kent, P.~Parekh, P.~Saab, P.~Balaji, P.~Rittner, P.~Bontrager, P.~Roux, P.~Dollar, P.~Zvyagina, P.~Ratanchandani, P.~Yuvraj, Q.~Liang, R.~Alao, R.~Rodriguez, R.~Ayub, R.~Murthy, R.~Nayani, R.~Mitra, R.~Li, R.~Hogan, R.~Battey, R.~Wang, R.~Maheswari, R.~Howes, R.~Rinott, S.~Bondu, Jayesh,
  S.~Datta, S.~Chugh, S.~Hunt, S.~Dhillon, S.~Sidorov, S.~Pan, S.~Verma, S.~Yamamoto, S.~Ramaswamy, S.~Lindsay, S.~Lindsay, S.~Feng, S.~Lin, S.~Zha, Cindy, S.~Shankar, S.~Zhang, S.~Zhang, S.~Wang, S.~Agarwal, S.~Sajuyigbe, S.~Chintala, S.~Max, S.~Chen, S.~Kehoe, S.~Satterfield, S.~Govindaprasad, S.~Gupta, S.~Cho, S.~Virk, S.~Subramanian, S.~Choudhury, S.~Goldman, T.~Remez, T.~Glaser, T.~Best, T.~Kohler, T.~Robinson, T.~Li, T.~Zhang, T.~Matthews, T.~Chou, T.~Shaked, V.~Vontimitta, V.~Ajayi, V.~Montanez, V.~Mohan, V.~Kumar, Satish, V.~Mangla, V.~Albiero, V.~Ionescu, V.~Poenaru, V.~Mihailescu, Tiberiu, V.~Ivanov, W.~Li, W.~Wang, W.~Jiang, W.~Bouaziz, W.~Constable, X.~Tang, X.~Wang, X.~Wu, X.~Wang, X.~Xia, X.~Wu, X.~Gao, Y.~Chen, Y.~Hu, Y.~Jia, Y.~Qi, Y.~Li, Y.~Zhang, Y.~Zhang, Y.~Adi, Y.~Nam, Yu, Wang, Y.~Hao, Y.~Qian, Y.~He, Z.~Rait, Z.~DeVito, Z.~Rosnbrick, Z.~Wen, Z.~Yang, and Z.~Zhao, ``{The Llama 3 Herd of Models},'' 2024.

\bibitem{socc-2024-aws}
X.~Fu, Z.~Zhang, H.~Fan, G.~Huang, M.~El-Shabani, R.~Huang, R.~Solanki, F.~Wu, R.~Diamant, and Y.~Wang, ``{Distributed Training of Large Language Models on AWS Trainium},'' in \emph{Proceedings of the 2024 ACM Symposium on Cloud Computing (SoCC)}, 2024.

\bibitem{iedm-2020-lpddr-hybrid-bonding}
B.~Fujun, J.~Xiping, W.~Song, Y.~Bing, T.~Jie, Z.~Fengguo, W.~Chunjuan, W.~Fan, L.~Xiaodong, Y.~Guoqing, F.~Ni, L.~Qiannan, L.~Hua, W.~Kexin, D.~Huifu, B.~Liang, J.~Xuerong, L.~Jin, L.~Mei, W.~Zhengwen, H.~Sheng, Z.~Jun, Z.~Qiong, S.~Peng, Y.~Daohong, C.~Kau, D.~Yang, C.-S. Ho, S.~Hongbin, L.~Hangbing, L.~Ming, K.~Yi, and R.~Qiwei, ``{A Stacked Embedded DRAM Array for LPDDR4/4X using Hybrid Bonding 3D Integration with 34GB/s/1Gb 0.88pJ/b Logic-to-Memory Interface},'' in \emph{IEEE International Electron Devices Meeting (IEDM)}, 2020.

\bibitem{google-2025-tpuv7}
\BIBentryALTinterwordspacing
Google, ``{Ironwood: The First Google TPU for the Age of Inference},'' 2025. [Online]. Available: \url{https://blog.google/products/google-cloud/ironwood-tpu-age-of-inference/}
\BIBentrySTDinterwordspacing

\bibitem{micro-2016-improving-energh-efficiency}
H.~Ha, A.~Pedram, S.~Richardson, S.~Kvatinsky, and M.~Horowitz, ``{Improving Energy Efficiency of DRAM by Exploiting Half Page Row Access},'' in \emph{MICRO}, 2016.

\bibitem{micro-2024-smd}
H.~Hassan, A.~Olgun, A.~G. Yağlıkçı, H.~Luo, O.~Mutlu, and E.~Zurich, ``{Self-Managing DRAM: A Low-Cost Framework for Enabling Autonomous and Efficient DRAM Maintenance Operations},'' in \emph{MICRO}, 2024.

\bibitem{asplos-2024-neupims}
G.~Heo, S.~Lee, J.~Cho, H.~Choi, S.~Lee, H.~Ham, G.~Kim, D.~Mahajan, and J.~Park, ``{NeuPIMs: NPU-PIM Heterogeneous Acceleration for Batched LLM Inferencing},'' in \emph{ASPLOS}, 2024.

\bibitem{intel-2024-pagepolicy}
Intel, ``{Performance Differences for Open-Page/Close-Page Policy},'' \url{https://www.intel.com/content/www/us/en/content-details/826015/performance-differences-for-open-page-close-page-policy.html}, 2024.

\bibitem{sec-2024-zenhammer}
P.~Jattke, M.~Wipfli, F.~Solt, M.~Marazzi, M.~B{\"o}lcskei, and K.~Razavi, ``{ZenHammer: Rowhammer Attacks on AMD Zen-based Platforms},'' in \emph{USENIX Security Symposium}, 2024.

\bibitem{jedec-2013-hbm}
JEDEC, ``{High Bandwidth Memory (HBM) DRAM},'' 2013.

\bibitem{jedec-2017-ddr4}
JEDEC, ``{DDR4 SDRAM Standard},'' 2017.

\bibitem{jedec-2018-hbm2}
JEDEC, ``{High Bandwidth Memory DRAM (HBM1, HBM2) Standard},'' 2018.

\bibitem{jedec-2022-hbm3}
JEDEC, ``{High Bandwidth Memory DRAM (HBM3) Standard},'' 2022.

\bibitem{jedec-2024-ddr5}
JEDEC, ``{DDR5 SDRAM Standard},'' 2024.

\bibitem{jedec-2025-hbm4}
JEDEC, ``{High Bandwidth Memory (HBM4) DRAM},'' 2025.

\bibitem{isca-2023-tpuv4}
N.~Jouppi, G.~Kurian, S.~Li, P.~Ma, R.~Nagarajan, L.~Nai, N.~Patil, S.~Subramanian, A.~Swing, B.~Towles, C.~Young, X.~Zhou, Z.~Zhou, and D.~A. Patterson, ``{TPU v4: An Optically Reconfigurable Supercomputer for Machine Learning with Hardware Support for Embeddings},'' in \emph{ISCA}, 2023.

\bibitem{micro-2011-minimalist}
D.~Kaseridis, J.~Stuecheli, and L.~K. John, ``{Minimalist Open-page: A DRAM Page-mode Scheduling Policy for the Many-core Era},'' in \emph{MICRO}, 2011.

\bibitem{isca-2024-tcp}
H.~Kim, Y.~Choi, J.~Park, B.~Bae, H.~Jeong, S.~M. Lee, J.~Yeon, M.~Kim, C.~Park, B.~Gu, C.~Lee, J.~Bae, S.~Bae, Y.~Cha, W.~Choe, J.~Choi, J.~Ha, H.~Han, N.~Hwang, S.~Hwang, K.~Jang, H.~Je, H.~Jeon, J.~Jeon, H.~Jeong, Y.~Jung, D.~Kang, H.~Kim, M.~Kim, M.~Kim, S.~Kim, S.~Kim, W.~Kim, Y.~Kim, Y.~Kim, Y.~Ku, J.~K. Lee, J.~Lee, K.~Lee, S.~Lee, M.~Noh, H.~Oh, G.~Park, S.~Park, J.~Seo, J.~Seong, J.~Paik, N.~P. Lopes, and S.~Yoo, ``{TCP: A Tensor Contraction Processor for AI Workloads},'' in \emph{ISCA}, 2024.

\bibitem{hpca-2010-atlas}
Y.~Kim, D.~Han, O.~Mutlu, and M.~Harchol-Balter, ``{ATLAS: A Scalable and High-Performance Scheduling Algorithm for Multiple Memory Controllers},'' in \emph{HPCA}, 2010.

\bibitem{micro-2010-TCM}
Y.~Kim, M.~Papamichael, O.~Mutlu, and M.~Harchol-Balter, ``{Thread Cluster Memory Scheduling: Exploiting Differences in Memory Access Behavior},'' in \emph{HPCA}, 2010.

\bibitem{isscc-2014-hbm}
D.~U. Lee, K.~W. Kim, K.~W. Kim, H.~Kim, J.~Y. Kim, Y.~J. Park, J.~H. Kim, D.~S. Kim, H.~B. Park, J.~W. Shin, J.~H. Cho, K.~H. Kwon, M.~J. Kim, J.~Lee, K.~W. Park, B.~Chung, and S.~Hong, ``{25.2 A 1.2V 8Gb 8-Channel 128GB/s High-Bandwidth Memory (HBM) Stacked DRAM with Effective Microbump I/O Test Methods Using 29nm Process and TSV},'' in \emph{IEEE International Solid-State Circuits Conference Digest of Technical Papers (ISSCC)}, 2014.

\bibitem{isscc-2024-hbm3e}
J.~Lee, K.~Cho, C.~K. Lee, Y.~Lee, J.-H. Park, S.-H. Oh, Y.~Ju, C.~Jeong, H.~S. Cho, J.~Lee, T.-S. Yun, J.~H. Cho, S.~Oh, J.~Moon, Y.-J. Park, H.-S. Choi, I.-K. Kim, S.~M. Yang, S.-Y. Kim, J.~Jang, J.~Kim, S.-H. Lee, Y.~Jeon, J.~Park, T.-K. Kim, D.~Ka, S.~Oh, J.~Kim, J.~Jeon, S.~Kim, K.~T. Kim, T.~Kim, H.~Yang, D.~Yang, M.~Lee, H.~Song, D.~Jang, J.~Shin, H.~Kim, C.~Baek, H.~Jeong, J.~Yoon, S.-K. Lim, K.~Y. Lee, Y.~J. Koo, M.-J. Park, J.~Cho, and J.~Kim, ``{13.4 A 48GB 16-High 1280GB/s HBM3E DRAM with All-Around Power TSV and a 6-Phase RDQS Scheme for TSV Area Optimization},'' in \emph{IEEE International Solid-State Circuits Conference (ISSCC)}, 2024.

\bibitem{hpca-2017-partial-row}
Y.~Lee, H.~Kim, S.~Hong, and S.~Kim, ``{Partial Row Activation for Low-Power DRAM System},'' in \emph{HPCA}, 2017.

\bibitem{sc-2012-mage}
S.~Li, D.~H. Yoon, K.~Chen, J.~Zhao, J.~Ahn, J.~B. Brockman, Y.~Xie, and N.~P. Jouppi, ``{MAGE: Adaptive Granularity and ECC for Resilient and Power Efficient Memory Systems},'' in \emph{SC}, 2012.

\bibitem{arxiv-2024-deepseek-v2}
\BIBentryALTinterwordspacing
A.~Liu, B.~Feng, B.~Wang, B.~Wang, B.~Liu, C.~Zhao, C.~Dengr, C.~Ruan, D.~Dai, D.~Guo, D.~Yang, D.~Chen, D.~Ji, E.~Li, F.~Lin, F.~Luo, G.~Hao, G.~Chen, G.~Li, H.~Zhang, H.~Xu, H.~Yang, H.~Zhang, H.~Ding, H.~Xin, H.~Gao, H.~Li, H.~Qu, J.~Cai, J.~Liang, J.~Guo, J.~Ni, J.~Li, J.~Chen, J.~Yuan, J.~Qiu, J.~Song, K.~Dong, K.~Gao, K.~Guan, L.~Wang, L.~Zhang, L.~Xu, L.~Xia, L.~Zhao, L.~Zhang, M.~Li, M.~Wang, M.~Zhang, M.~Zhang, M.~Tang, M.~Li, N.~Tian, P.~Huang, P.~Wang, P.~Zhang, Q.~Zhu, Q.~Chen, Q.~Du, R.~Chen, R.~Jin, R.~Ge, R.~Pan, R.~Xu, R.~Chen, S.~Li, S.~Lu, S.~Zhou, S.~Chen, S.~Wu, S.~Ye, S.~Ma, S.~Wang, S.~Zhou, S.~Yu, S.~Zhou, S.~Zheng, T.~Wang, T.~Pei, T.~Yuan, T.~Sun, W.~Xiao, W.~Zeng, W.~An, W.~Liu, W.~Liang, W.~Gao, W.~Zhang, X.~Li, X.~Jin, X.~Wang, X.~Bi, X.~Liu, X.~Wang, X.~Shen, X.~Chen, X.~Chen, X.~Nie, X.~Sun, X.~Wang, X.~Liu, X.~Xie, X.~Yu, X.~Song, X.~Zhou, X.~Yang, X.~Lu, X.~Su, Y.~Wu, Y.~Li, Y.~Wei, Y.~Zhu, Y.~Xu, Y.~Huang, Y.~Li, Y.~Zhao, Y.~Sun, Y.~Li, Y.~Wang, Y.~Zheng, Y.~Zhang, Y.~Xiong,
  Y.~Zhao, Y.~He, Y.~Tang, Y.~Piao, Y.~Dong, Y.~Tan, Y.~Liu, Y.~Wang, Y.~Guo, Y.~Zhu, Y.~Wang, Y.~Zou, Y.~Zha, Y.~Ma, Y.~Yan, Y.~You, Y.~Liu, Z.~Ren, Z.~Ren, Z.~Sha, Z.~Fu, Z.~Huang, Z.~Zhang, Z.~Xie, Z.~Hao, Z.~Shao, Z.~Wen, Z.~Xu, Z.~Zhang, Z.~Li, Z.~Wang, Z.~Gu, Z.~Li, and Z.~Xie, ``{DeepSeek-V2: A Strong, Economical, and Efficient Mixture-of-Experts Language Model},'' 2024. [Online]. Available: \url{https://arxiv.org/abs/2405.04434}
\BIBentrySTDinterwordspacing

\bibitem{arxiv-2023-ramulator2}
\BIBentryALTinterwordspacing
H.~Luo, Y.~C. Tuğrul, F.~N. Bostancı, A.~Olgun, A.~G. Yağlıkçı, and O.~Mutlu, ``{Ramulator 2.0: A Modern, Modular, and Extensible DRAM Simulator},'' 2023. [Online]. Available: \url{https://github.com/CMU-SAFARI/ramulator2.git}
\BIBentrySTDinterwordspacing

\bibitem{github-2025-trainpolicy}
\BIBentryALTinterwordspacing
MLPerf, ``{Training Policies},'' accessed in Oct 2025. [Online]. Available: \url{https://github.com/mlcommons/training_policies/tree/master}
\BIBentrySTDinterwordspacing

\bibitem{micro-2007-stfm}
O.~Mutlu and T.~Moscibroda, ``{Stall-Time Fair Memory Access Scheduling for Chip Multiprocessors},'' in \emph{MICRO}, 2007, pp. 146--160.

\bibitem{isca-2008-parbs}
O.~Mutlu and T.~Moscibroda, ``{Parallelism-Aware Batch Scheduling: Enhancing both Performance and Fairness of Shared DRAM Systems},'' in \emph{ISCA}, 2008.

\bibitem{isca-2024-dramscope}
H.~Nam, S.~Baek, M.~Wi, M.~J. Kim, J.~Park, C.~Song, N.~S. Kim, and J.~Ahn, ``{DRAMScope: Uncovering DRAM Microarchitecture and Characteristics by Issuing Memory Commands},'' in \emph{{ISCA}}, 2024.

\bibitem{sc-2021-microbatch}
D.~Narayanan, M.~Shoeybi, J.~Casper, P.~LeGresley, M.~Patwary, V.~Korthikanti, D.~Vainbrand, P.~Kashinkunti, J.~Bernauer, B.~Catanzaro, A.~Phanishayee, and M.~Zaharia, ``{Efficient Large-Scale Language Model Training on GPU Clusters Using Megatron-LM},'' in \emph{SC}, 2021.

\bibitem{micro-2006-fair-queuing}
K.~J. Nesbit, N.~Aggarwal, J.~Laudon, and J.~E. Smith, ``{Fair Queuing Memory Systems},'' in \emph{MICRO}, 2006, pp. 208--222.

\bibitem{isscc-2022-logic-hybrid-bonding}
D.~Niu, S.~Li, Y.~Wang, W.~Han, Z.~Zhang, Y.~Guan, T.~Guan, F.~Sun, F.~Xue, L.~Duan, Y.~Fang, H.~Zheng, X.~Jiang, S.~Wang, F.~Zuo, Y.~Wang, B.~Yu, Q.~Ren, and Y.~Xie, ``{184QPS/W 64Mb/mm23D Logic-to-DRAM Hybrid Bonding with Process-Near-Memory Engine for Recommendation System},'' in \emph{{IEEE International Solid-State Circuits Conference (ISSCC)}}, 2022.

\bibitem{nvidia-2020-a100}
\BIBentryALTinterwordspacing
NVIDIA, ``{NVIDIA A100 Tensor Core GPU Architecture},'' 2020. [Online]. Available: \url{https://images.nvidia.com/aem-dam/en-zz/Solutions/data-center/nvidia-ampere-architecture-whitepaper.pdf}
\BIBentrySTDinterwordspacing

\bibitem{nvidia-h100}
\BIBentryALTinterwordspacing
NVIDIA, ``{NVIDIA H100 GPU},'' 2024. [Online]. Available: \url{https://resources.nvidia.com/en-us-hopper-architecture/nvidia-tensor-core-gpu-datasheet}
\BIBentrySTDinterwordspacing

\bibitem{nvidia-b200}
\BIBentryALTinterwordspacing
NVIDIA, ``{NVIDIA Blackwell Architecture Technical Brief},'' 2025. [Online]. Available: \url{https://resources.nvidia.com/en-us-blackwell-architecture}
\BIBentrySTDinterwordspacing

\bibitem{nvidia-ai-factory}
\BIBentryALTinterwordspacing
NVIDIA, ``{NVIDIA Blackwell Architecture Technical Brief},'' 2025. [Online]. Available: \url{https://www.nvidia.com/en-us/solutions/ai-factories}
\BIBentrySTDinterwordspacing

\bibitem{isca-2014-rbd}
S.~O, Y.~H. Son, N.~S. Kim, and J.~Ahn, ``Row-buffer decoupling: A case for low-latency dram microarchitecture,'' in \emph{ISCA}, 2014.

\bibitem{micro-2017-fgdram}
M.~O'Connor, N.~Chatterjee, D.~Lee, J.~Wilson, A.~Agrawal, S.~W. Keckler, and W.~J. Dally, ``{Fine-Grained DRAM: Energy-Efficient DRAM for Extreme Bandwidth Systems},'' in \emph{MICRO}, 2017.

\bibitem{isscc-2020-hbm2e}
C.-S. Oh, K.~C. Chun, Y.-Y. Byun, Y.-K. Kim, S.-Y. Kim, Y.~Ryu, J.~Park, S.~Kim, S.~Cha, D.~Shin, J.~Lee, J.-P. Son, B.-K. Ho, S.-J. Cho, B.~Kil, S.~Ahn, B.~Lim, Y.~Park, K.~Lee, M.-K. Lee, S.~Baek, J.~Noh, J.-W. Lee, S.~Lee, S.~Kim, B.~Lim, S.-K. Choi, J.-G. Kim, H.-I. Choi, H.-J. Kwon, J.~J. Kong, K.~Sohn, N.~S. Kim, K.-I. Park, and J.-B. Lee, ``{22.1 A 1.1V 16GB 640GB/s HBM2E DRAM with a Data-Bus Window-Extension Technique and a Synergetic On-Die ECC Scheme},'' in \emph{IEEE International Solid-State Circuits Conference (ISSCC)}, 2020.

\bibitem{taco-2024-sectored-dram}
A.~Olgun, F.~N. Bostanci, G.~Francisco~de Oliveira~Junior, Y.~C. Tugrul, R.~Bera, A.~G. Yaglikci, H.~Hassan, O.~Ergin, and O.~Mutlu, ``{Sectored DRAM: A Practical Energy-Efficient and High-Performance Fine-Grained DRAM Architecture},'' \emph{TACO}, 2024.

\bibitem{arxiv-2025-gptoss}
\BIBentryALTinterwordspacing
OpenAI, ``{gpt-oss-120b \& gpt-oss-20b Model Card},'' 2025. [Online]. Available: \url{https://arxiv.org/abs/2508.10925}
\BIBentrySTDinterwordspacing

\bibitem{asplos-2024-attacc}
J.~Park, J.~Choi, K.~Kyung, M.~J. Kim, Y.~Kwon, N.~S. Kim, and J.~Ahn, ``{AttAcc! Unleashing the Power of PIM for Batched Transformer-based Generative Model Inference},'' in \emph{ASPLOS}, 2024.

\bibitem{isscc-2022-hbm3}
M.-J. Park, H.~S. Cho, T.-S. Yun, S.~Byeon, Y.~J. Koo, S.~Yoon, D.~U. Lee, S.~Choi, J.~Park, J.~Lee, K.~Cho, J.~Moon, B.-K. Yoon, Y.-J. Park, S.-m. Oh, C.~K. Lee, T.-K. Kim, S.-H. Lee, H.-W. Kim, Y.~Ju, S.-K. Lim, S.~G. Baek, K.~Y. Lee, S.~H. Lee, W.~S. We, S.~Kim, Y.~Choi, S.-H. Lee, S.~M. Yang, G.~Lee, I.-K. Kim, Y.~Jeon, J.-H. Park, J.~C. Yun, C.~Park, S.-Y. Kim, S.~Kim, D.-Y. Lee, S.-H. Oh, T.~Hwang, J.~Shin, Y.~Lee, H.~Kim, J.~Lee, Y.~Hur, S.~Lee, J.~Jang, J.~Chun, and J.~Cho, ``{A 192-Gb 12-High 896-GB/s HBM3 DRAM with a TSV Auto-Calibration Scheme and Machine-Learning-Based Layout Optimization},'' in \emph{IEEE International Solid-State Circuits Conference (ISSCC)}, 2022.

\bibitem{isca-2024-splitwise}
P.~Patel, E.~Choukse, C.~Zhang, A.~Shah, Ã.~Goiri, S.~Maleki, and R.~Bianchini, ``{Splitwise: Efficient Generative LLM Inference Using Phase Splitting},'' in \emph{ISCA}, 2024.

\bibitem{sec-2016-drama}
P.~Pessl, D.~Gruss, C.~Maurice, M.~Schwarz, and S.~Mangard, ``{DRAMA: Exploiting DRAM Addressing for Cross-CPU Attacks},'' in \emph{25th USENIX Security Symposium (USENIX Security 16)}, 2016.

\bibitem{micro-2013-locality-aware}
M.~Rhu, M.~Sullivan, J.~Leng, and M.~Erez, ``{A Locality-Aware Memory Hierarchy for Energy-Efficient GPU Architectures},'' in \emph{MICRO}, 2013.

\bibitem{isca-2000-fr-fcfs}
S.~Rixner, W.~J. Dally, U.~J. Kapasi, P.~Mattson, and J.~D. Owens, ``{Memory Access Scheduling},'' in \emph{ISCA}, 2000.

\bibitem{jssc-2023-samsung-hbm3}
Y.~Ryu, S.-G. Ahn, J.~H. Lee, J.~Park, Y.~K. Kim, H.~Kim, Y.~G. Song, H.-W. Cho, S.~Cho, S.~H. Song, H.~Lee, U.~Shin, J.~Ahn, J.-M. Ryu, S.~Lee, K.-H. Lim, J.~Lee, J.~H. Park, J.-S. Jeong, S.~Joo, D.~Cho, S.~Y. Kim, M.~Lee, H.~Kim, M.~Kim, J.-S. Kim, J.~Kim, H.~G. Kang, M.-K. Lee, S.-R. Kim, Y.-C. Kwon, Y.~Y. Byun, K.~Lee, S.~Park, J.~Youn, M.-O. Kim, K.~Sohn, S.-J. Hwang, and J.~Lee, ``{A 16 GB 1024 GB/s HBM3 DRAM With Source-Synchronized Bus Design and On-Die Error Control Scheme for Enhanced RAS Features},'' \emph{IEEE Journal of Solid-State Circuits}, 2023.

\bibitem{SKhynix-tsv-pitch}
\BIBentryALTinterwordspacing
{SK hynix}, ``{Advanced Packaging Technology for Beyond Memory},'' 2023. [Online]. Available: \url{https://www.theise.org/wp-content/uploads/2023/10/Tutorial1-4_%EC%86%90%ED%98%B8%EC%98%81%EC%88%98%EC%84%9D%EB%8B%98_SK%ED%95%98%EC%9D%B4%EB%8B%89%EC%8A%A4.pdf}
\BIBentrySTDinterwordspacing

\bibitem{isca-2024-amd}
A.~Smith, G.~H. Loh, M.~J. Schulte, M.~Ignatowski, S.~Naffziger, M.~Mantor, M.~Fowler, N.~Kalyanasundharam, V.~Alla, N.~Malaya, J.~L. Greathouse, E.~Chapman, and R.~Swaminathan, ``{Realizing the AMD Exascale Heterogeneous Processor Vision},'' in \emph{ISCA}, 2024.

\bibitem{sc-2014-microbank}
Y.~H. Son, O.~Seongil, H.~Yang, D.~Jung, J.~Ahn, J.~Kim, J.~Kim, and J.~W. Lee, ``{Microbank: Architecting Through-Silicon Interposer-Based Main Memory Systems},'' in \emph{SC}, 2014.

\bibitem{asplos-2010-micro-pages}
K.~Sudan, N.~Chatterjee, D.~Nellans, M.~Awasthi, R.~Balasubramonian, and A.~Davis, ``{Micro-pages: Increasing DRAM Efficiency with Locality-aware Data Placement},'' in \emph{ASPLOS}, 2010.

\bibitem{hynix_base_die_process}
\BIBentryALTinterwordspacing
{The Korea Economic Daily Global Edition}, ``{SK Hynix Ships World's First 12-Layer HBM4 Samples Early},'' 2025. [Online]. Available: \url{https://www.kedglobal.com/korean-chipmakers/newsView/ked202503190006}
\BIBentrySTDinterwordspacing

\bibitem{samsung_base_die_process}
\BIBentryALTinterwordspacing
{The Korean Economic Daily}, ``{Samsung to Mass-Produce HBM4 on 4 nm Foundry Process},'' 2024. [Online]. Available: \url{https://www.kedglobal.com/korean-chipmakers/newsView/ked202407150016}
\BIBentrySTDinterwordspacing

\bibitem{arxiv-2023-llama2}
\BIBentryALTinterwordspacing
H.~Touvron, L.~Martin, K.~Stone, P.~Albert, A.~Almahairi, Y.~Babaei, N.~Bashlykov, S.~Batra, P.~Bhargava, S.~Bhosale \emph{et~al.}, ``{Llama 2: Open Foundation and Fine-Tuned Chat Models},'' 2023. [Online]. Available: \url{https://arxiv.org/abs/2307.09288}
\BIBentrySTDinterwordspacing

\bibitem{isca-2010-rethinking}
A.~N. Udipi, N.~Muralimanohar, N.~Chatterjee, R.~Balasubramonian, A.~Davis, and N.~P. Jouppi, ``{Rethinking DRAM Design and Organization for Energy-Constrained Multi-Cores},'' in \emph{ISCA}, 2010.

\bibitem{neurips-2017-transformer}
A.~Vaswani, N.~Shazeer, N.~Parmar, J.~Uszkoreit, L.~Jones, A.~N. Gomez, L.~u. Kaiser, and I.~Polosukhin, ``{Attention is All You Need},'' in \emph{NeurIPS}, 2017.

\bibitem{dac-2020-dramdig}
M.~Wang, Z.~Zhang, Y.~Cheng, and S.~Nepal, ``{DRAMDig: A Knowledge-assisted Tool to Uncover DRAM Address Mapping},'' in \emph{Design Automation Conference (DAC)}, 2020.

\bibitem{dramsec-2025-sudoku}
M.~Wi, S.~Baek, S.~Park, M.~Erez, and J.~H. Ahn, ``{Sudoku: Decomposing DRAM Address Mapping into Component Functions},'' \emph{arXiv preprint arXiv:2506.15918}, 2025.

\bibitem{github-2024-grok1}
\BIBentryALTinterwordspacing
{xAI}, ``{grok1},'' {2024}. [Online]. Available: \url{https://github.com/xai-org/grok-1}
\BIBentrySTDinterwordspacing

\bibitem{isca-2011-adaptive-granularity}
D.~H. Yoon, M.~K. Jeong, and M.~Erez, ``{Adaptive Granularity Memory Systems: A Tradeoff between Storage Efficiency and Throughput},'' in \emph{ISCA}, 2011.

\bibitem{isca-2012-dyanmic-granularity}
D.~H. Yoon, M.~K. Jeong, M.~Sullivan, and M.~Erez, ``{The Dynamic Granularity Memory System},'' in \emph{ISCA}, 2012.

\bibitem{osdi-2022-orca}
G.-I. Yu, J.~S. Jeong, G.-W. Kim, S.~Kim, and B.-G. Chun, ``{Orca: A Distributed Serving System for {Transformer-Based} Generative Models},'' in \emph{USENIX Symposium on Operating Systems Design and Implementation (OSDI 22)}, 2022.

\bibitem{micro-2024-duplex}
S.~Yun, K.~Kyung, J.~Cho, J.~Choi, J.~Kim, B.~Kim, S.~Lee, K.~Sohn, and J.~Ahn, ``{Duplex: A Device for Large Language Models with Mixture of Experts, Grouped Query Attention, and Continuous Batching},'' in \emph{MICRO}, 2024.

\bibitem{arxiv-2025-mla}
\BIBentryALTinterwordspacing
S.~Yun, S.~Park, H.~Nam, Y.~Lee, G.~Lee, K.~Kyung, S.~Kim, N.~S. Kim, J.~Kim, H.~Kim, J.~Cho, S.~Baek, and J.~Ahn, ``{The New LLM Bottleneck: A Systems Perspective on Latent Attention and Mixture-of-Experts},'' 2025. [Online]. Available: \url{https://arxiv.org/abs/2507.15465}
\BIBentrySTDinterwordspacing

\bibitem{islpled-2017-enabling-efficient-finegrained}
C.~Zhang and X.~Guo, ``{Enabling Efficient Fine-Grained DRAM Activations with Interleaved I/O},'' in \emph{ISLPED}, 2017.

\bibitem{isca-2014-half-dram}
T.~Zhang, K.~Chen, C.~Xu, G.~Sun, T.~Wang, and Y.~Xie, ``{Half-DRAM: A High-bandwidth and Low-power DRAM Architecture from the Rethinking of Fine-grained Activation},'' in \emph{ISCA}, 2014.

\bibitem{iccd-2023-morpheus}
X.~Zhang, T.~Lu, Y.~Chang, K.~Zhang, and M.~Chen, ``{Morpheus: An Adaptive DRAM Cache with Online Granularity Adjustment for Disaggregated Memory},'' in \emph{International Conference on Computer Design (ICCD)}, 2023.

\bibitem{micro-2000-interleaving}
Z.~Zhang, Z.~Zhu, and X.~Zhang, ``{A Permutation-based Page Interleaving Scheme to Reduce Row-buffer Conflicts and Exploit Data Locality},'' in \emph{MICRO}, 2000.

\bibitem{micro-2008-minirank}
H.~Zheng, J.~Lin, Z.~Zhang, E.~Gorbatov, H.~David, and Z.~Zhu, ``{Mini-Rank: Adaptive DRAM Architecture for Improving Memory Power Efficiency},'' in \emph{MICRO}, 2008.

\end{thebibliography}

\end{document}